\documentclass[english,prc,showpacs,amsmath,amssymb, floatfix,nofootinbib, preprint]{revtex4}
\usepackage[T1]{fontenc}
\usepackage[latin1]{inputenc}
\usepackage{graphicx}% Include figure files
\makeatletter

\usepackage{array}
\usepackage{longtable}
\usepackage{babel}
\usepackage{subfigure}

\def\beq{\begin{equation}}
\def\eeq{\end{equation}}
\def\ba{\begin{eqnarray}}
\def\ea{\end{eqnarray}}

\def\v8p{v_8^\prime}

\newcommand{\boldsigma}{\mbox{\boldmath$\sigma$}}
\newcommand{\boldtau}{\mbox{\boldmath$\tau$}}

\newcommand{\sij}{\boldsigma_i \cdot \boldsigma_j}
\newcommand{\isij}{\boldtau_i \cdot \boldtau_j}
\newcommand{\lds}{{\bf L} \cdot {\bf S}}
\newcommand{\tij}{S_{ij}}

\newcommand{\bec}{\begin{center}}
\newcommand{\enc}{\end{center}}
\newcommand{\bit}{\begin{itemize}}
\newcommand{\eit}{\end{itemize}}

\newcommand{\bfk}{{\bf k}}

\newcommand{\bfq}{{\bf q}}
\newcommand{\bfr}{{\bf r}}
\newcommand{\bft}{{\bf t}}

\newcommand{\calF}{{\cal F}}

\newcommand{\calO}{{\cal O}}

\newcommand{\calT}{{\cal T}}

\usepackage{babel}
\makeatother

\begin{document}

\title{Variational Theory of Hot Nucleon Matter II : Spin-Isospin Correlations and Equation of State of Nuclear and Neutron Matter}

\author{Abhishek Mukherjee}
\email{amukherj@illinois.edu}

\affiliation{Department of Physics, University of Illinois at
  Urbana-Champaign,
  \\
  1110 W. Green St., Urbana, IL 61801, U.S.A.}

\date{\today{}}

\begin{abstract}
  We apply the variational theory for fermions at finite temperature and high
  density, developed in an earlier paper, to
  symmetric nuclear matter and pure neutron matter. This extension generalizes
  to finite temperatures, the many body technique used in the
  construction of the zero temperature Akmal-Pandharipande-Ravenhall
  equation of state. We discuss how the formalism can be used for
  practical calculations of hot dense matter. Neutral pion
  condensation along with the associated isovector spin longitudinal
  sum rule is analyzed. The equation of state is calculated
  for temperatures less than 30 MeV and densities less than three times the
  saturation density of nuclear matter. The behavior of the nucleon effective
  mass in medium is also discussed.    
\end{abstract}


\maketitle

\section{Introduction}

\textit{Ab initio} models of dense nuclear matter at finite temperature are crucial to the
understanding of supernovae evolution \cite{LatPra2000, Jan2007}, composition \cite{Pra1997} and cooling \cite{Pon1999} of  protoneutron stars, gravitational wave emission spectrum from neutron star mergers \cite{OecJan2007}, 
and the analysis of heavy ion collision experiments \cite{LiCheKo2008}. Nuclear matter at
zero temperature has been studied extensively using a variety of
theoretical methods (see, e.g., \cite{HeiPan2000} for a review). In contrast, the corresponding many body problem at finite temperature has received little attention \cite{Bal1995}. For example, most computer simulations
of supernovae explosions use phenomenological equations of state,
like the Lattimer-Swesty equation of state \cite{LatSwe1991}
or the equation of state due to Shen et al. \cite {She1998npa,She1998ptp}. 

Three decades ago Friedman and Pandharipande (FP) carried out a seminal calculation of the equation of state of hot dense nuclear and neutron matter using a variational theory
\cite{FriPan1981}. Since then only a few other calculations have been carried out. The methods employed include Bloch-de Dominicis diagrammatic expansion
\cite{BalFer1999}, extended Brueckner theory \cite{Lej1986, Zuo2004},
self consistent Green's functions \cite{Rio2006}, Dirac-Brueckner
theory \cite{TerMal1986}, relativistic
Brueckner-Hartree-Fock theory \cite{HubWebWei1998}, perturbation theory with low momentum interactions \cite{TolFriSch2008}, and  lowest order
variational \cite{Kan2007} methods.

Most variational theories of dense quantum fluids originate from  Jastrow's  original suggestion that a
reasonable approximation for the
wave functions  ($\Psi$) of an interacting system can be obtained by writing them
as a product of the wave functions of a noninteracting system ($\Phi$) and a
product of pair correlation functions ($F_{ij}$) \cite{Jas1955},
\beq
\label{jas}
\Psi \varpropto \prod_{i<j} F_{ij} \Phi \ .
\eeq 
The pair correlation functions are found  by minimizing the ground state energy at zero temperature or the free energy at finite temperature. Different generalizations of this basic variational method have been applied to the  many body problem
with varying degree of success and sophistication \cite{Cla1979}. In the approach used here the simple pair correlation function in Eq.~(\ref{jas}) is replaced by a pair correlation operator to include the effects of non-central correlations directly in the variational wave functions.  The calculation of the energy expectation values are subsequently more difficult. Nevertheless a
reasonably accurate calculation can be done using a combination of the Fermi hypernetted chain  summation and the single operator chain summation approximations \cite{PanWir1979rmp} (and references therein). The combination of the Fermi hypernetted chain summation method for central correlations, the single operator chain summation for non-central correlations and some improvements introduced in later works (see next paragraph) is collectively known as the variational chain summation method \cite{AkmPan1997}.

The equation of state of dense nucleonic matter at zero temperature has been
calculated using the variational chain summation method by FP, then by Wiringa, Fiks and Fabrocini
\cite{WirFikFab1988} and by Akmal, Pandharipande and Ravenhall (APR)
\cite{AkmPan1997, AkmPanRav1998}. Each successive calculation
was a significant improvement over the  preceding one
with respect to the variational method, and the models of the nucleon-nucleon 
interaction and three nucleon interaction used.
Recently the three body cluster in the variational chain summation method was 
calculated exactly (VCS/3)  \cite{MorPanRav2002} unlike earlier calculations
including APR where only the two body cluster terms were calculated exactly and all the  higher
order terms were calculated approximately (VCS/2). However VCS/3 has
not yet been generalized to finite temperatures nor is a full zero
temperature equation of state available at this moment. Thus in this work we have
used the VCS/2 method and henceforth we will refer to VCS/2 simply as variational chain summation.

We believe that variational chain summation supplemented by realistic nucleon-nucleon interactions and three nucleon interactions is at present one of the most reliable
methods for studying dense many body systems. It compares reasonably well
with the experimental bounds on the equation of state set by heavy ion collision
experiments \cite{DanLacLyn2002} and benchmark quantum Monte Carlo
calculations in neutron matter \cite{CarMorPanRav2003}. It is, therefore,
worthwhile to investigate its generalization to nonzero temperatures
and arbitrary proton fraction. 

In extending variational chain summation to finite temperatures one is faced with a technical challenge. Since wave functions used in variational chain summation are 
not mutually orthogonal, when constructing a variational theory at
finite temperature one encounters the so called orthogonality
corrections \cite{SchPan1979, FanPan1988}. The orthogonality
corrections are not unique (they depend on the method of
orthonormalization chosen) and  more importantly their calculation
requires the evaluation of the off-diagonal matrix elements of the
Hamiltonian and unity. At present there exists no accurate method to calculate these off-diagonal matrix elements, especially for wave functions with operator dependent correlations. 

In the only previous application of the variational method to finite temperatures,
viz. in FP, the orthogonality corrections were simply ignored. Recently we showed that there
exists a choice for the orthonormalization procedure such that the
orthogonality corrections to the free energy  vanish in the
thermodynamic limit \cite {MukPan2007} (hereafter I). This way the
orthogonality problem can be circumvented and at least for thermodynamic
quantities the variational chain summation method can be extended to nonzero temperatures without having to worry about orthogonality corrections.    

In this paper we apply the formalism developed in I to symmetric
nuclear matter  and pure neutron matter. The present calculations can be regarded as an improvement on the finite temperature calculations due to  FP  and/or an attempt to generalize  the zero temperature calculations of APR to finite temperature. 
This work is a contriubtion to the ongoing effort to understand the properties  of nucleonic matter under extreme conditions of density, temperature and asymmetry starting from the bare interactions amongst nucleons which can be constrained by the experimental data on the nucleon-nucleon scattering in vacuum and the binding energies of light nuclei.

 The Hamiltonian we have used for this
 calculation is the same as that in APR. It consists of the Argonne $v18$ nucleon-nucleon interaction \cite{Wir1995v18}, the Urbana IX
 model of three nucleon interaction  \cite{CarPanWir1983NNN, Pud1995uix} and leading
 order relativistic boost interactions \cite{ForPanFri1995}. For completeness we will discuss the Hamiltonian in some detail in the next section. In
 Section~\ref{sec:vcs} we outline the variational chain summation method for finite temperature. First a
 short summary of the relevant parts of I is given. Thereafter the
 practicalities of the search for the variational minima are described. Section~\ref{sec:res} is devoted to our results. We first
 discuss the spin-isospin correlations in symmetric nuclear matter and pure neutron matter, and show that
 they are enhanced at finite temperature as they are at zero
 temperature \cite{AkmPan1997}. The fate of the neutral pion
 condensation at finite temperature is discussed. The behavior of
 the nucleon effective mass, free energy, pressure and symmetry energy
 is also discussed in this section. Some of the limitations of our
 calculations are discussed in Section~\ref{sec:dis}. We summarize our results in
 Section~\ref{sec:con}.

\section{The Hamiltonian}
\label{sec:ham}

The non-relativistic Hamiltonian used in the present calculations can
be written in real space as

\beq 
H= \sum_i -\frac{\hbar^2}{4}\left[ \left( \frac{1}{m_p} +
    \frac{1}{m_n} \right) + \left( \frac{1}{m_p} - \frac{1}{m_n}
  \right) \tau_{zi} \right] \nabla_i^2 + \sum_{ij} \left ( v_{18,ij} +
  \delta v_{b,ij} \right) + \sum_{ijk}
V^{\star}_{\mbox{\tiny{UIX}},ijk} % \\
% &=& K + \sum_{ij} \left ( v_{18,ij} +\delta v_{b,ij}
% \right)+\sum_{ijk} V_{ijk}^{\star}
\eeq 
where the kinetic energy operator takes into account
the difference between the mass of a proton $m_p$ and the mass of a
neutron $m_n$.

The Argonne $v_{18}$ nucleon-nucleon interaction $v_{18,ij}$ has the form 
\beq 
v_{18,ij} =
\sum_{p=1}^{18} v^p_{ij} O^p_{ij}+v_{em} \ .  
\eeq 
The electromagnetic
part $v_{em}$ is omitted from all nuclear matter studies including
the present one. The strong interaction part has fourteen isoscalar
operator terms 
\beq 
O^{p=1,14}_{ij}=\left [1, \sij,\tij, \left ( \lds
  \right )_{ij},L_{ij}^2,L^2_{ij}\sij, \left ( \lds \right
  )^2_{ij}\right ]\otimes \left [1, \isij \right ] \ .  
\eeq 
By convention the operators with even $p \leq 14$ have the $\isij$ term
while the ones with odd $p$ do not. The three isotensor operators 
($p=15,16,17$) are given by 
\beq
O^{p=15,18}_{ij}=(3\tau_{zi}\tau_{zj}-\isij)\otimes (1,\sij,S_{ij}) \ .  
\eeq 
And finally, the isovector operator ($p=18$) is 
\beq
O^{18}_{ij} = (\tau_{zi}+\tau_{zj}) \ .  
\eeq 
The Argonne $v_{18}$ nucleon-nucleon interaction
along with the CD BONN \cite{MacSamSon1996cdbonn, Mac2001cdbonn} and the Nijmegen models
\cite{Sto1994nim} constitute the set of `modern' phase-shift equivalent
nucleon-nucleon interactions. These models fit the Nijmegen data base of proton-proton and
neutron-proton scattering phase shifts up to $350$ MeV with a
$\chi^2/N_{\mbox{\tiny{data}}}\sim 1$. All of them include the long
range one pion interaction potential but have different treatments of
the intermediate and short range parts of the nucleon-nucleon interaction.  

The isotensor and isovector parts of $v_{18,ij}$, and the isovector part
of the kinetic energy, are very weak and we will treat them
as first order perturbations. In first order, these terms do not
contribute to the energy of symmetric nuclear matter, which has total isospin $\calT=0$. In pure neutron matter the
isovector and the isotensor terms can be absorbed in the central, spin and the tensor parts of the nucleon-nucleon interaction.

The Urbana IX model of three nucleon interaction $ V_{\mbox{\tiny{UIX}},ijk}$ has two terms 
\beq
V_{\mbox{\tiny{UIX}},ijk}=V^{2\pi}_{ijk} + V^{R}_{ijk} \ .  
\eeq 
The first term represents the Fujita-Miyazawa two-pion exchange
interaction 
\ba 
&V^{2\pi}_{ijk}=\sum_{cyc} A_{2\pi} \left(
  \left\{\isij, \boldtau_i \cdot \boldtau_k \right\} \left\{
    X_{ij},X_{ik} \right\} + \frac{1}{4} [\isij, \boldtau_i \cdot
  \boldtau_k]
  [X_{ij},X_{ik}] \right) \ ,&  \\
&X_{ij} = S_{ij}T_\pi(r_{ij})+\sij Y_\pi(r_{ij}) \ ,& 
\ea
with strength $A_{2\pi}$. The functions $T_{\pi}(r_{ij})$ and
$Y_{\pi}(r_{ij})$ describe the radial shapes of the one-pion exchange
tensor and Yukawa potentials. The term denoted by $V^R_{ijk}$ is
purely phenomenological, and has the form 
\beq
V^{R}_{ijk}=U_0\sum_{cyc}T_\pi^2(r_{ij})T_\pi^2(r_{ik}).  
\eeq 
This term is meant to represent the modification of N$\Delta$- and
$\Delta\Delta$-contributions in the two-body interaction by other
particles in the medium. The two parameters $A_{2\pi}$ and $U_0$ are
chosen to yield the observed energy of $^3$H and the equilibrium
density of cold symmetric nuclear matter, $\rho_0=0.16$ fm$^{-3}$.

It is possible  to incorporate the leading order effects of relativistic boost
corrections as an interaction term  into a
non-relativistic Hamiltonian \cite{ForPanFri1995}. They give rise to two body interactions
$\delta v_{ij}$. As a first approximation, only the static part of
the boost interaction is kept in the present calculation. Studies of light nuclei using the
Variational Monte Carlo method \cite{CarPanSch1993} find that the contribution of the two-body
boost interaction to the energy is repulsive, with a magnitude which
is 37\% of the $V^R_{ijk}$ contribution. However, the parameters of
the Urbana IX model of three nucleon interaction were fixed without taking into account the
boost interactions.  In the  new model of three nucleon interaction called UIX$^{\star}$ the parameters are obtained by fitting the binding energies of $^3$H and $^4$He, and the equilibrium density of symmetric nuclear matter,
including $\delta v_{ij}$.  The strength of $V^{R\ast}_{ijk}$ is 0.63 times
that of $V^R_{ijk}$ in UIX, while  $V^{2 \pi}_{ijk}$ remains unchanged \cite{AkmPanRav1998}.

\section{Variational Chain Summation at Finite Temperature}
\label{sec:vcs}
\subsection{The variational theory at finite temperature}
In variational calculations one assumes that a good approximation for
the eigenstates of the interacting system of fermions is given by the
correlated basis states  \cite{SchPan1979, FriPan1981}
\beq 
\Psi_i \left [ n_i(\bfk,x)\right ]= \frac{\mathcal{S}\left ( \prod_{i<j} \calF_{ij} \right ) \Phi_i \left [
  n_i(\bfk,x)\right ]}{\Phi_i \left [
  n_i(\bfk,x)\right ] \left (\mathcal{S}\left ( \prod_{i<j} \calF_{ij}
  \right )\right )^2 \Phi_i \left [
  n_i(\bfk,x)\right ]} \label{wavefn} 
\eeq 
where $\Phi_i$ are the eigenstates of a free Fermi gas with occupation numbers $n_i\left (
  \bfk,x \right )$ for single particle states with momentum $\bfk$ and
$x$ standing for any other quantum numbers of the single particle
states (including spin and isospin). The occupation numbers $n_i\left
  (\bfk,x \right )$ can take the values $0$ and $1$.  The pair correlation
operator $\calF_{ij}$ encodes the effects of interactions. In the present
calculations it has the following form 
\beq 
\calF_{ij} = \sum_{i=1}^8 f^p \left( r_{ij} \right ) O^p_{ij} \ .
\eeq 
Since the operators $O^p_{ij}$ do not commute, the product of pair correlation operators
need to be symmetrized with the symmetrization operator $\mathcal{S}$
to make the full wave function antisymmetric. Henceforth we will denote
the static operator channels $p=1-6$ interchangeably with the more
descriptive symbols $c,c\tau,\sigma,\sigma\tau,t$ and $t\tau$ which
stand for central, central-isospin, spin, spin-isospin, tensor and
tensor-isospin respectively. 
 
The ansatz Eq.~(\ref{wavefn}) is consistent with Landau's theory of Fermi liquids where it
is assumed that the eigenstates of an interacting system have a one-to-one correspondence with the eigenstates of a non-interacting system. In this approach we assume that the mapping is accomplished by
the correlation operator. In the spirit of Landau's theory we will
hereafter refer to the $n_i(\bfk,x)$ as the quasiparticle occupation numbers.

An upper bound for the free energy of an interacting system can be
obtained by using the Gibbs-Bogoliubov variational principle \cite{FeyStatMech}
\beq 
F \leq F_v = \mbox{Tr} \left ( \rho_v H \right) + T \mbox{Tr}
\left ( \rho_v \ln \rho_v \right ) \label{varpri} \ , 
\eeq
 where $T$ is the temperature of the system and $\rho_v$ is a
 variational density matrix (not to be confused with the density
 $\rho$). The inequality is replaced by an equality when $\rho_v$ is the
exact density matrix of the system. Since we assume that the correlated basis states
provide a good approximation to the eigenstates of the interacting
system we want to construct $\rho_v$ from the correlated basis states. However, the correlated basis states, by
construction, are not mutually orthonormal, i.e. in general  
\beq
\langle \Psi_i | \Psi_j \rangle \ne 0 \mbox{   for } i \ne j \ .
\eeq
Hence, we need to
orthonormalize them before they can be used\footnote{It is possible to define the
  variational density matrix with the non-orthogonal correlated basis states. But in that case, the trace
  operation will involve their dual vectors and the free energy expectation
value will still contain off-diagonal matrix elements.} \cite{FanPan1988}. The
orthonormalization procedure is not unique. Let $\left |
\Theta_i \rangle \right .$ be the orthonormalized correlated basis states
 using one such orthonormalization method. Now, we 
can choose the following form for the variational density
matrix, 
\ba
&\rho_v = {\exp \left ( -\beta H_v \right) }/{\mbox{Tr}~ \ [ \exp\left(-\beta
    H_v \right ) }\ ] & \label{varden}\\
&H_v = E^v_i \left | \Theta_i \rangle \langle \Theta_i
\right | \ , & \label{varham} 
\ea 
where $\beta=1/T$ 

and the variational spectrum can be
approximated as a sum of quasiparticle energies \cite{SchPan1979} 
\beq
E^v_i \left [n \left ( \bfk, x \right ) \right ] = \sum_{\bfk,x}
\epsilon \left (\bfk,x;\rho,T \right ) n\left (\bfk, x \right ) \
. \label{quasieng} 
\eeq 
The quasiparticle energies $\epsilon \left (\bfk,x;\rho,T \right )$
depend on the density $\rho$ and the temperature $T$ along with $\bfk$ and $x$. 

The entropy is given by
\ba
  S_v(\rho,T) &=& -\mbox{Tr} \left ( \rho_v \ln \rho_v \right )\nonumber \\
         &=& - \sum \left [ \bar{n}(\bfk,x;\rho,T) \ln \lbrace \bar{n}(\bfk,x;\rho,T)  \rbrace \right . \nonumber\\
         & &  \left .  + \lbrace 1-\bar{n}(\bfk,x;\rho,T) \rbrace \ln \lbrace 1 -
     \bar{n}(\bfk,x;\rho,T) \rbrace \right ] \ . \label{entpy} 
\ea
where  $\bar{n}(\bfk,x;\rho,T)$ is the mean quasiparticle occupation number of the
single particle state $(\bfk,x)$ at density $\rho$ and temperature $T$,
\beq 
\bar{n}(\bfk,x;\rho,T)=\frac{1}{\exp \left [\beta
    \lbrace \epsilon \left ( \bfk,x;\rho,T \right )-\tilde\mu \left ( \rho, T \right ) \rbrace \right ]+1} \label{quasioccu} \ . 
\eeq
Here $\tilde{\mu}$ is an effective chemical potential (see next subsection) and is fixed by 
\beq 
A=\sum_{\bfk,x}\bar{n}(\bfk,x;\rho, T) \ , \label{quasimu} 
\eeq 
where $A$ is the total number of particles in the system. We have set the Boltzmann constant $k_B$ to 1.

The thermodynamic average of the Hamiltonian is given by
\ba
\mbox{Tr} \left ( \rho_v H \right)&=& \langle H
\rangle_{\mbox{\tiny{OCBS}},\bar{n}(\bfk,x;\rho,T)} \\
                                  &=& \langle H \rangle_{\mbox{\tiny{CBS}},\bar{n}(\bfk,x;\rho,T)}
                                    + E_{\mbox{\tiny{OC}}}(\rho,T) \label{expocbs}
\ea
where  $\langle H \rangle_{\mbox{\tiny{CBS}},\bar{n}(\bfk,x;\rho,T)} $  and $\langle H \rangle_{\mbox{\tiny{OCBS}},\bar{n}(\bfk,x;\rho,T)}$ are  the expectation values of the Hamiltonian in the
  correlated basis states and orthonormalized correlated basis states, respectively, with the occupation numbers set to $\bar{n} \left (\bfk,x;\rho,T \right )$. The orthogonality corrections are denoted by $E_{\mbox{\tiny{OC}}}$.       

The calculation of the expectation values of various operators,
especially the Hamiltonian, in the basis of correlated basis states is a nontrivial
problem. The diagonal matrix elements and thus the expectation value
in Eq.~(\ref{expocbs}) can be calculated by expanding
it in powers of $\calF^2_{ij}-1$. Schematically we have
\ba
 E_v(\rho,T)/A &=& \langle H
 \rangle_{\mbox{\tiny{CBS}},\bar{n}(\bfk,x;\rho,T)}/A \nonumber \\
&=& \frac{\hbar^2}{2m} k_{\textrm{av}}^2 + \sum \textrm{diagrams} (v,\mathcal{F},l_{T}) \label{expcbs}
\ea

where $m$ is the average bare mass of a nucleon and $k_{\textrm{av}}^2$ is the mean square momentum per particle.
The diagrams are many body integrals involving the potential $v$, the
correlation operator $\mathcal{F}$ and the finite temperature Slater
function $l_{T} $,

\beq 
l_{T}(r ; \rho, T) = \frac{1}{A}\sum_{\bfk,x} \mbox{e}^{i\mbox{\bfk} \cdot \mbox{\bfr}}
\bar{n}(\bfk,x;\rho,T) \ . \label{slaterT}
\eeq 
Within our scheme the pair correlation operator $\calF$ and the Slater function
$l_{T}(r;\rho,T)$ encode all the relevant microscopic information about
the many particle system at a temperature $T$ and density
$\rho$.

 Large classes of diagrams can be resummed using the
variational chain summation method. For a review of the variational chain summation
method the reader is referred to \cite{PanWir1979rmp, AkmPan1997, WirFikFab1988, AkmPanRav1998} (and references
therein).  Here,  we merely wish to point out that the variational chain summation method can be used to
calculate the expectation values for any set of mean occupation
numbers $\bar{n} \left (\bfk,x \right )$. At zero temperature this set is a step
function, but at finite temperatures states above the Fermi
surface become populated.

In contrast to $E_v(\rho,T)$ whose calculation involves only the
diagonal matrix elements of the Hamiltonian in the correlated basis states basis, the orthogonality corrections
$E_{\mbox{\tiny{OC}}}$ involve off-diagonal matrix elements of the
Hamiltonian and the unit operator in the correlated basis states
basis and they \emph{cannot} be calculated within variational chain summation. In fact, there 
exists \emph{no method} to calculate the off-diagonal matrix
elements accurately for operator dependent interactions and
correlation functions like the ones used in this study.

In FP it was assumed that these orthogonality corrections are
small. Recently it was shown in I that exploiting the fact that only a very small subset of states
contributes to  the trace in Eq.~(\ref{varpri}), it is possible to construct an
orthonormalization scheme where the orthogonality corrections
to the free energy vanish in the thermodynamic limit  
\beq
\frac{E_{\mbox{\tiny{OC}}}}{A} \to 0 \mbox{ as } A
\to \infty \ \label{ortho}.
\eeq
This proof does not rely on
the detailed nature of the pair correlation functions, instead it follows
from some general properties that any reasonable pair correlation function
(including ours) is expected to possess.    

Hence using Eqs.~(\ref{entpy}, \ref{expocbs}, \ref{expcbs}, \ref{ortho}) it
is possible to rewrite Eq.~(\ref{varpri}) exactly as, 
\beq 
F(\rho,T)<F_v(\rho,T)  =  E_v(\rho,T) - T
S_v(\rho,T) \label{varquasi02} %\\
%\implies f(\rho,T)<f_v(\rho,T)  &=& e_v(\rho,T)- T s_v(\rho,T) 
%\langle H \rangle_{\mbox{\tiny{Ncorrelated basis states}},
%  n(\bfk,x)=\bar{n}(\bfk,x)} \nonumber \\
%& & + T \sum_{\bfk,x} \left [ \bar{n}(\bfk,x) \ln \bar{n}(\bfk,x) +
%  \left ( 1-\barn(\bfk,x) \right ) \ln \left ( 1-\barn(\bfk,x) \right ) \right ] \label{varquasi} \\
%=  \langle H \rangle_{\mbox{\tiny{correlated basis states}},\barn(\bfk,x;\rho,T) - T S_v(\rho,T) \label{varquasi02} 
\eeq 
%where $f(\rho,T)\left ( =F(\rho,T)/A \right )$ is the free energy per particle of the system and $f_v(\rho,T)$ is its variational upper bound. 
In the above equation  $S_v$ can be obtained trivially from Eq.(\ref{entpy}) once the single
particle spectrum is known and $E_v$ is calculated using variational chain summation
generalized to finite temperature as outlined above.
For further details the reader is referred to I.

\subsection{The Optimization Procedure and the Pair Correlation Functions}

The variational free energy  $F_v$ is optimized by varying both the
single particle spectrum $\epsilon(\bfk,x;\rho,T)$ and the correlation
operator $\calF_{ij}$ at all densities and temperatures. This should be contrasted with the
calculations of FP where for densities $\rho > 0.04$ fm$^{-3}$ the zero temperature
correlation operator was used at all temperatures. We find that the
temperature dependence of the correlation operator is weak but non-negligible (see later). 

We parametrize the  single particle spectrum $\epsilon(\bfk, x;\rho,T)$  by a
simple effective mass approximation 
\beq 
\epsilon(\bfk,x;\rho,T) =
\frac{\hbar^2 k^2}{ 2 m^{\star}(\rho,T)} \ .  
\eeq 
In general it is possible for $\epsilon(\bfk,x;\rho,T)$ to have higher order terms in $k$. However, in our calculations
$F_v$ was found to be insensitive to any such dependence. It is also
possible for $\epsilon(\bfk,x;\rho,T)$ to contain a momentum independent term
which depends on $\rho$ and $T$ only, $u(\rho,T)$, but such a term will be absorbed in
the definition of $\tilde{\mu}$ \cite{PanRav1989}
\beq
\tilde{\mu}(\rho,T)=\mu(\rho,T)-u(\rho,T)
\eeq 
where $\mu$ is the true chemical potential of the system.

As mentioned earlier, in variational chain summation the expectation values of various
operators are calculated by expanding the matrix elements in powers of
$\calF^2_{ij}-1$ and then resumming large classes of terms. The
correlation operator $\calF_{ij}$ is itself calculated by solving the
Euler-Lagrange equations which are obtained by minimizing the lowest
order terms ( viz., the sum of the kinetic energy term and the two body
cluster term) in this expansion for the Hamiltonian, with the nucleon-nucleon interaction
$v_{ij}$ replaced by $\bar{v}_{ij}-\lambda_{ij}$, where 
\ba
\bar{v}_{ij} & = & \sum_{p=1,14}\alpha^p v^p(r_{ij}) O^p_{ij}, \\
\lambda_{ij} & = & \sum_{p=1,8}\lambda^p(r_{ij})O^p_{ij}.  
\ea 
The variational parameters $\alpha^p$ are meant to simulate the quenching
of the spin-isospin interaction between particles $i$ and $j$, due to
flipping of the spin and/or isospin of particle $i$ or $j$ via interaction
with other particles in matter. We use 
\ba
\alpha^p & = & 1 \ \mbox{for} \  p=1 \ \mbox{and} \ 9  \\
\alpha^p & = & \alpha \ \mbox{otherwise} \ .  
\ea

The correlation functions $f^p(r)$ are made to satisfy the additional
healing conditions 
\ba
f^p(r>d^p) &=& \delta_{p1} \\
\left . \frac{df^p}{dr} \right |_{r=d^p}&=&0 
\ea 
where 
\ba
d^p & = & d_t \ \mbox{for} \ p=5,6 \\
d^p & = & d_c \ \mbox{for} \ p \neq 5,6.  
\ea 
The above constraints completely determine the $\lambda^p(r)$.

The Euler-Lagrange equations are solved in the spin $S$ and isospin $\calT$ channels. As an example, in the $S=0$ channel the Euler-Lagrange equations become 
\beq 
-\frac{\hbar^2}{2 m}\left [ \phi_{\calT,S=0}
  \nabla^2 f_{\calT, S=0} + 2 \nabla \phi_{\calT,S=0}\cdot \nabla
  f_{\calT,S=0}\right ] + \left ( v_{\calT,S=0}-\lambda_{\calT,S=0} \right )
f_{\calT,S=0} \phi_{\calT,S=0} =0  \label{el}
\eeq 
where 
\beq 
\phi_{\calT,S}=\left [ 1-(-1)^{(\calT+S)} l_{T}^2(r ; \rho,T) \right ]^{1/2} \ .
\eeq
 
The relationship between the potential and the correlation functions in
the $\calT,S$ channels and those in the operator channels $p=1-8$ are
given in \cite{PanWir1979rmp}. The Euler-Lagrange equations in the $S=1$
channels are considerably more complicated because the contribution of
the tensor and the spin-orbit terms give rise to three coupled
differential equations for the correlation functions in each isospin channel. For more details
the reader is referred to \cite{PanWir1979rmp}. 

\begin{figure}
\begin{center}
\includegraphics[width=\textwidth]{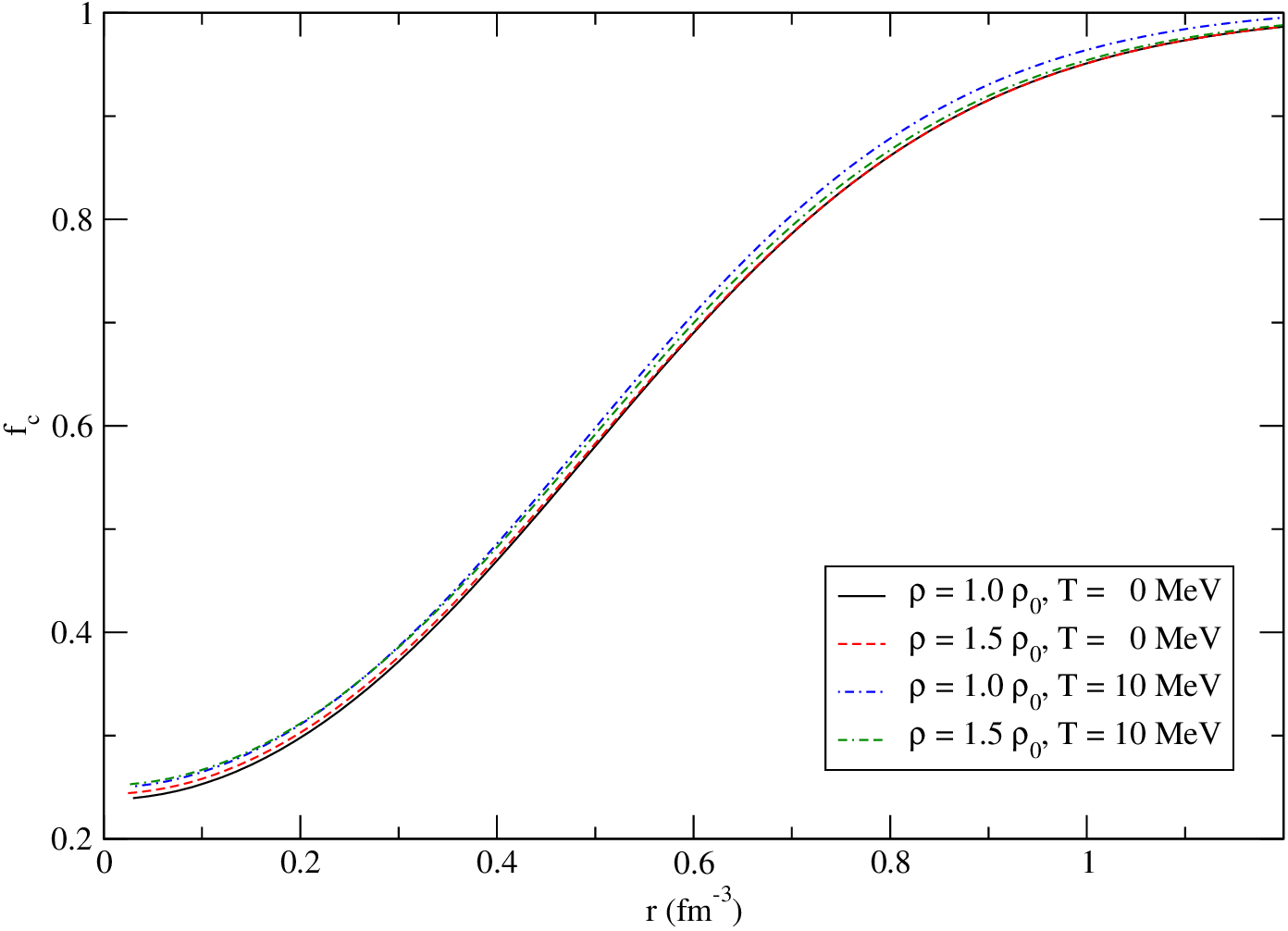}
\caption{ (Color online) The  pair correlation function in the central channel of symmetric nuclear matter
various densities and temperatures.}
\label{fig:fig1}
\end{center}
\end{figure}

\begin{figure}
\begin{center}
\includegraphics[width=\textwidth]{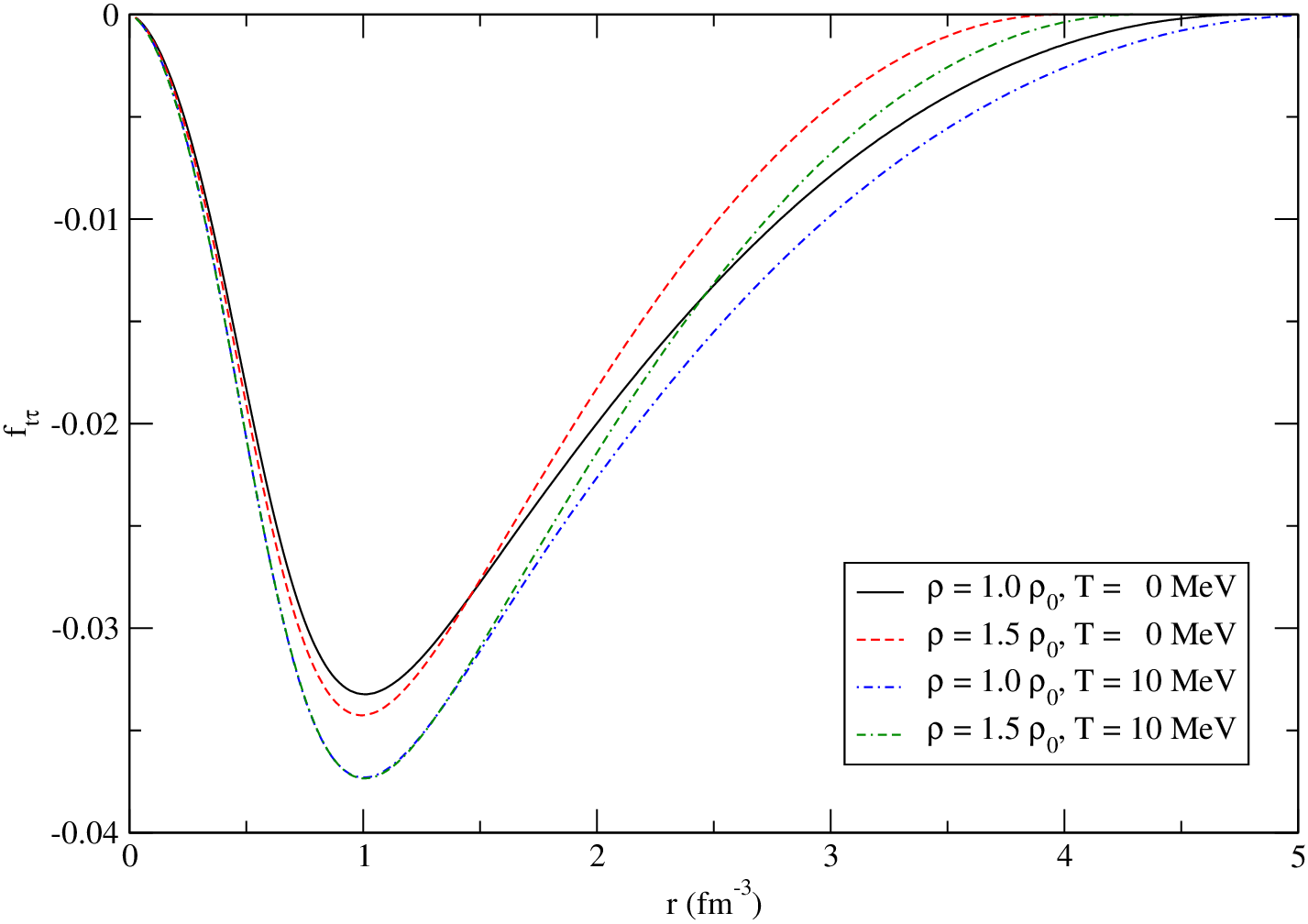}
\caption{ (Color online) The pair correlation function in the tensor-isospin channel  of symmetric nuclear matter
various densities and temperatures.}
\label{fig:fig2}
\end{center}
\end{figure}

To illustrate the relative effect of density and temperature on the 
pair correlation functions we show in Figs.~\ref{fig:fig1} and  \ref{fig:fig2} the pair correlation 
functions in the central ($f^c$) and the tensor-isospin ($f^{t\tau}$) 
channels for $\rho=\rho_0$ and $1.5 \rho_0$ and $T=0$  and $10$ MeV.   
As we mentioned in the beginning of this section, thermal effects
 (at least in the range of temperatures we are interested in) are not negligible but they do not change the behavior of the $f^p$'s qualitatively.   

The variational free energy $F_v(\rho,T)$ thus becomes a function of the four
variational parameters $m^{\star}, d_c, d_t$ and $\alpha$. The optimal values for
these parameters and hence the best value $F_v$ at each density and
temperature is found by minimizing the constrained free energy defined
as \cite{WirFikFab1988}
\beq 
F_{\mbox{\tiny{con}}} = F_v+ A \Lambda \left[ (I_c-1)^2 +
  (\frac{1}{3}I_\tau+1)^2 \right ] \label{fcon}
\eeq 
where 
\ba
I_c = \rho \int d^3r (1-g^c(r))  \\
I_{\tau} = \frac{1}{N} \langle 0| \sum_{i,j=1,N} \isij |0\rangle 
\ea
and $g^c(r)$ is the pair distribution function. Laws of conservation of
mass and charge demand that,
\ba
I_c &=& 1 \label{sumc}\\
I_{\tau} &=& -3 \label{sumtau}\ .  
\ea 
Of course, for pure neutron matter only Eq.~(\ref{sumc}) is applicable and hence, only the first term within square brackets in Eq.~(\ref{fcon}) is kept.

In the right hand side of Eq.~(\ref{fcon}) penalty term is added to
$F_v$ to make sure that these laws of conservation of mass and charge are
approximately satisfied in the variational calculations. In our
calculations we choose $\Lambda$ to be $1000$ MeV \cite{AkmPan1997}. This keeps $I_c$
and $I_{\tau}$ to within about $5\%$ of their exact values for all values of
$\rho$ and $T$. For most values of $\rho$ and $T$ the conservation
laws are satisfied at the level of one percent or less. 

The temperature dependence of the correlation operator is important in satisfying the sum rules Eqs.~(\ref{sumc}) and (\ref{sumtau}) to a reasonable degree of accuracy. For example, at $\rho = \rho_0$ and $T=10$ MeV, the deviation of $I_c$ and $I_{\tau}$ in our calculations from their true values (Eqs.~(\ref{sumc}) and (\ref{sumtau})) is $ \approx 0.04\%$ and $\approx 0.3\%$, respectively. For comparison,   employ the methodology used by FP, the so called `frozen correlation' method, where the correlation operator is taken from the zero temperature calculations, only $m^{\star}$ is varied at finite temperature, and the minimization is carried out with  $F_v$ only. At the same values of $\rho$ and $T$ the deviation of $I_c$ and $I_{\tau}$ from their correct values is  $\approx 13\%$ and $\approx 20\%$, respectively, in this case.

  In practice we varied $m^{\star}/m$, $d_t/r_0$,
  $d_c/r_0$ and $\alpha$ during the search. Here $r_0$, is the unit radius, defined such
  that 
\beq 
\frac{4}{3} \pi r_0^3 \rho = 1 \ .  
\eeq 
  Due to technical reasons involving the spacing on the grid on which the integrations are done in our computer
  program, $d_c/r_0$ is varied in fixed steps  while the other three
  parameters are allowed to vary continuously. In particular,
  
\beq 
l_c=l_t \frac{d_c}{d_t} 
\eeq 
  could only take integer values
  where $2 l_t$ is the size of the grid on which most of the
  integrations are carried out. In our calculations we set $l_t$ to be
  $64$ while this was $32$ in the calculations due to APR. We
  also made improvements in the way the integrations are done while
  calculating the chain functions. While in APR a simple midpoint Euler method
  was used, we employed Gaussian quadrature for the same purpose.

  The actual variational search for each value of $l_c$ is carried
  out using a downhill simplex routine \cite{NelMea1965}. Within the context of variational chain summation the simplex search algorithm for the search parameters was first implemented in Ref.~\cite{WirFikFab1988}. Finally, the best parameters and
  $F_{\mbox{\tiny{con}}} $ are found from a quadratic fit to the corresponding values at the three
  best values of $l_c$. This is an improvement over APR where a simple grid search was done in the parameter space. 

  Typically the changes in energy due
  to these improved numerics are small ($\approx 0.5$ MeV per particle at $\rho=\rho_0$ and $T=0$), but the short range parts of
  the chain functions become much more well behaved as a result of
  these improvements, i.e., the quality of the final wave functions
  obtained is much improved \footnote{Some of the corresponding
    improvements for the zero temperature version of the computer
    program were first carried out by Jaime Morales \cite{Mor2008}.}.

  \section{Results}
\label{sec:res}

  \subsection{Two Body Densities and Spin-isospin Correlations} 
  \label{subsec:den}
\begin{figure}
\begin{center}
\includegraphics[width=\textwidth]{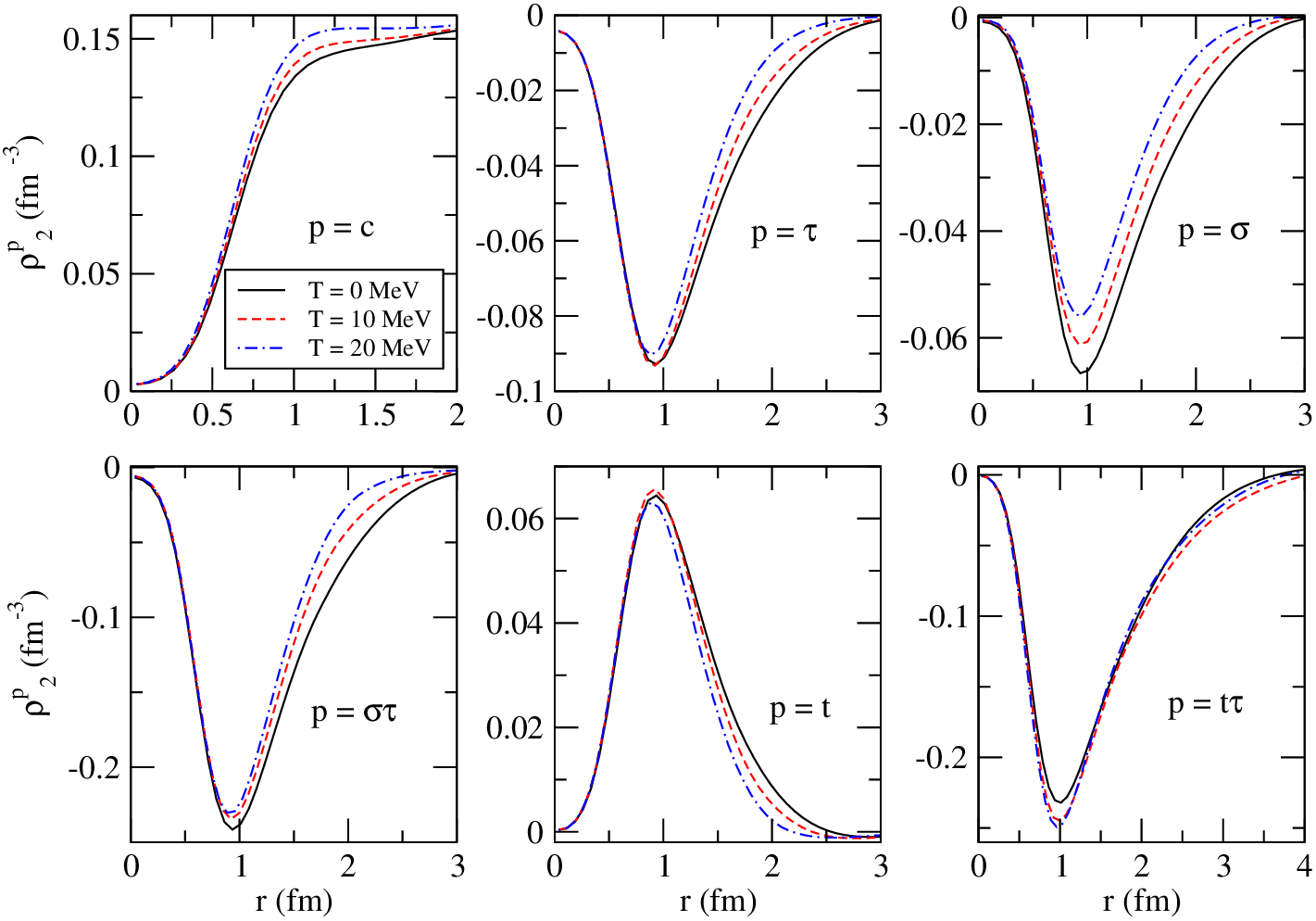}
\caption{ (Color online) The two body densities in symmetric nuclear matter at
  saturation density and  different temperatures.}
\label{fig:fig3}
\end{center}
\end{figure}

   The effect of nuclear interactions on two particle correlations in
   medium can be inferred from the two body densities $\rho^p_2(r)$,
   defined such that,
   \beq
   \langle \sum_{i \ne j=1,A}B(r_{ij})O^p_{ij} \rangle_{T,\rho} = A \int
   \mbox{d}^3 r B(r) \rho^p_2(r) \ ,
   \eeq    
   where $\langle \cdots \rangle_{T, \rho}$ denotes the  thermal average at
   temperature $T$ and density $\rho$. We show the static two body
   densities in Fig.~\ref{fig:fig3} for $T=0,10$ and $20$ MeV at
   saturation density $\rho=\rho_0$ in symmetric nuclear matter. Unsurprisingly, they have the
   expected asymptotic behavior
\beq
\rho^p_2(r \to \infty) = \rho \delta_{1p}.
\eeq
    The central two body density $\rho^c_2$  is particularly interesting because it is  $\rho$ times the probability of finding two particles separated by a distance $r$. In a non-interacting gas all non-central ($p>1$) two body
    densities vanish; their large magnitude in nuclear matter is due
    to strong spin-isospin correlations introduced by the nuclear interaction.
    
    The effect of increasing temperature on the two body densities, within the context of our calculations, is twofold. Firstly, increasing temperature weakens the effects of Pauli blocking in the Euler-Lagrange equations (like Eq.~(\ref{el})) for the pair correlation functions . This, tends to enhance short range correlations among nucleons. On the other hand, when a nonzero temperature is introduced the integral equations used in variational chain summation to calculate the the two body densities from the pair correlation functions, it tends to suppress the spin-isospin correlations. In the regime of temperatures and densities discussed in this work, the magnitudes of the two effects are comparable and the net temperature dependence of the two body densities is a combination of the two. 

\begin{figure}
\begin{center}
\includegraphics[width=\textwidth]{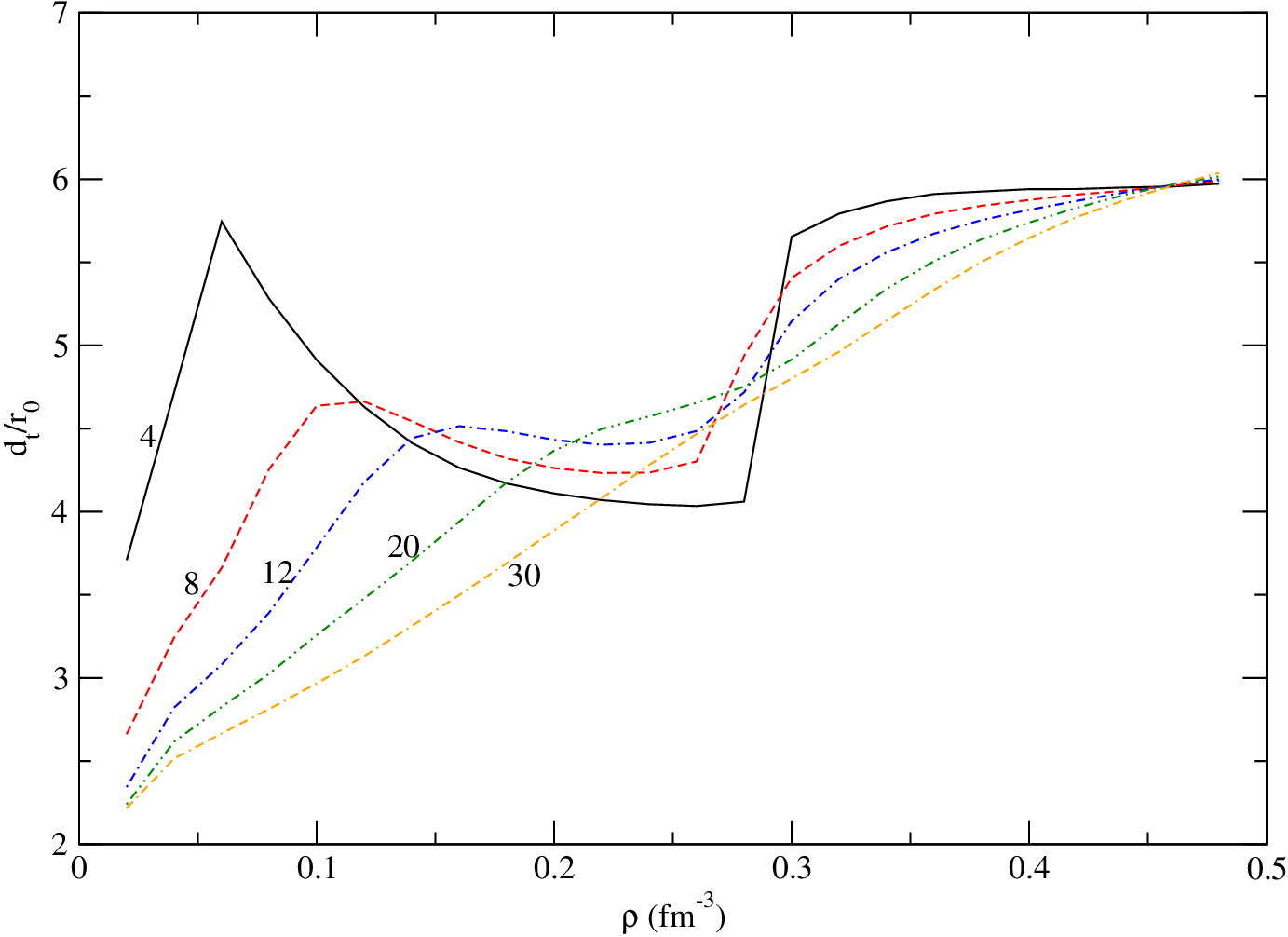}
\caption{ (Color online) The tensor correlation length in symmetric nuclear
  matter. The numbers alongside the curves denote the temperature in MeV. }
\label{fig:fig4}
\end{center}
\end{figure}

\begin{figure}
\begin{center}
\includegraphics[width=\textwidth]{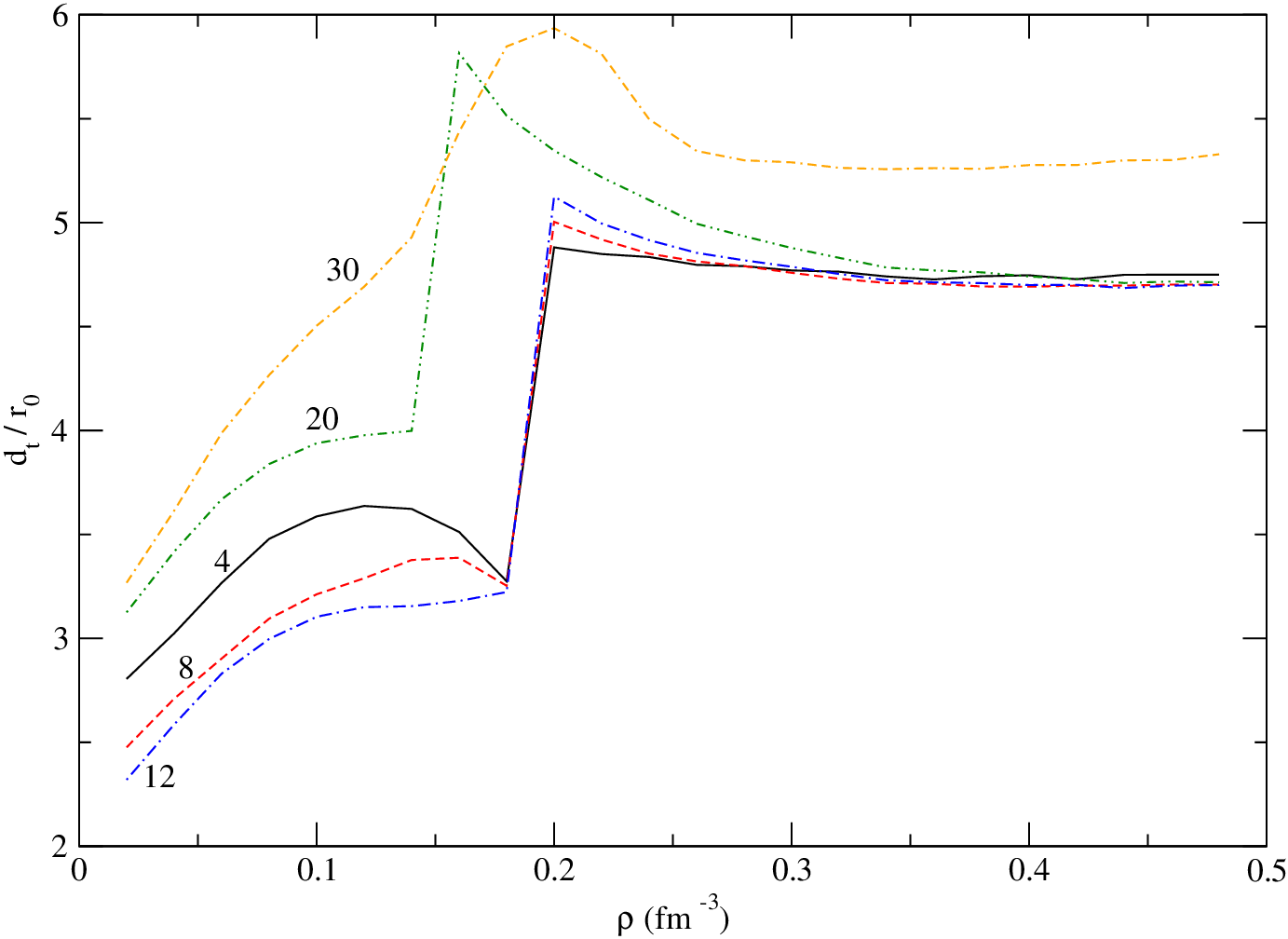}
\caption{ (Color online) The tensor correlation length in pure neutron matter. The numbers alongside the curves denote the temperature in MeV.}
\label{fig:fig5}
\end{center}
\end{figure}

 Generally particles are correlated over the longest range due to tensor interactions. In our calculations $d_t$  is one measure of this range. In Figs.~\ref{fig:fig4} and \ref{fig:fig5} we show $d_t$ for symmetric nuclear matter and pure neutron matter, respectively, for various temperatures. In Fig.~\ref{fig:fig4} we see a dramatic decrease in the value of $d_t/r_0$ below $\rho \sim 0.06$ fm$^{-3}$ for $T=4$ MeV in symmetric nuclear matter. This is a precursor to cluster formation. We will discuss this topic briefly in Section~\ref{sec:dis}. 

One of the most interesting features of these plots is the sharp change in magnitude of $d_t/r_0$ at $\rho
  \sim 2\rho_0$ for symmetric nuclear matter and at $\rho \sim \rho_0$ for pure neutron matter.
 A similar feature was obtained in the zero temperature calculations of APR. In APR it was argued that this feature resulted from a first order phase transition due to neutral pion condensation \cite{Mig1978, SawSca1973}. Their argument was based on the fact that  the energy per particle as a function of density showed a kink indicating a first order phase transition, and that the isovector spin longitudinal reponse showed an enhancement and softening in the high density phase compared to the low density phase indicating enhanced pion exchange interactions between nucleons in the high density phase. Recently there has been some indication of experimental  evidence supporting the enhancement of pionic modes in nuclei (see \cite{IchSakWak2006} for a review).

  The isovector spin longitudinal static structure function is defined as
\beq
S_L(\bfq) = \frac{1}{A} \left (\langle O_L^2 \rangle_{T,\rho} - |\langle O_L \rangle_{T,\rho}|^2\right ) \ ,
\eeq 
where
\beq
 O_L(\bfq) = \sum_{i=1}^A \boldsigma_i \cdot \hat{\bfq} \boldtau_i
 \cdot \hat{\bft} \mbox{e}^{i \bfq \cdot \bfr_i} \ . 
\eeq
Here $\bfr_i$ are the positions of the nucleons and $\hat{\bft}$ is a
unit vector in the isospin space. This quantity is also the sum of the
isovector spin longitudinal dynamic response function $R_L(\bfq,\omega)$,
\ba
&S_L(q)= \frac{1}{A}\int_0^\infty R_L(\bfq,\omega) \mbox{d}\omega& \\
&R_L(\bfq,\omega) = \sum_{I,J}^{} p_I \left |\langle J |O_L(\bfq)|I \rangle
  \right |^2 \delta(\omega_J-\omega_I - \omega) &
\ea 
where $I$ and $J$ are eigenstates of the Hamiltonian with eigenvalues
$\omega_I$ and $\omega_J$, respectively, and $p_I$ is the probability of
the state $I$ occurring in the thermal ensemble.

It is also possible to define a mean energy $\bar{E}_L(q)$ of the isovector spin longitudinal response  by,
\ba
W_L(q)= \frac{1}{A}\int_0^\infty \omega R_L(q,\omega)\textrm{d} \omega
\\
\bar{E}_L(q) = \frac{W_L(q)}{S_L(q)} 
\ea
 Both $S_L(q)$ and $W_L(q)$ and hence $\bar{E}_L(q)$ can be calculated from the
 two body densities $\rho^p_2(r)$.

\begin{figure}
\begin{center}
\includegraphics[width=\textwidth]{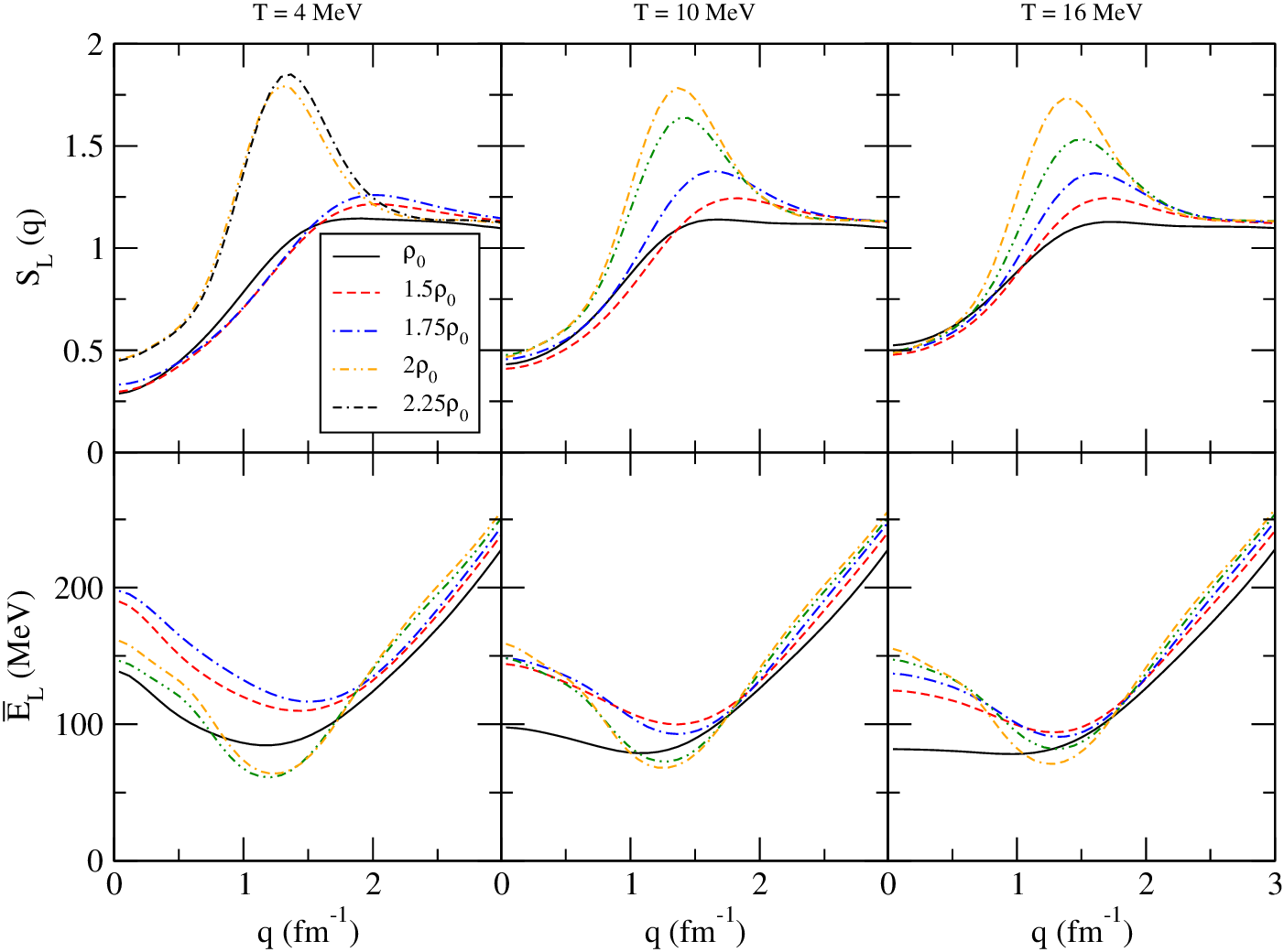}
\caption{ (Color online) The isovector spin longitudinal static structure function and mean energy in symmetric nuclear matter.}
\label{fig:fig6}
\end{center}
\end{figure}

  In Fig.~\ref{fig:fig6}  we plot $S_L$ and
  $\bar{E}_L$ in symmetric nuclear matter  at $T = 4 $, $10$  and $16$ MeV
  for various densities. For all three temperatures we see a big enhancement
  of $S_L$ at $q \sim 1.3$ fm$^{-1}$ due to stronger spin-isospin
  correlations at higher densities. However, for $T = 4$ MeV the
   enhancement develops quite suddenly at $\rho = 2\rho_0$
	 while at $T = 10$ MeV and $16$ MeV there is a smoother evolution of the enhancement as
  the density is increased. Similarly the mean energy $\bar{E}_L$ develops a dip at $q \sim 1.3$
   fm$^{-1}$ for all three temperatures, but quite sharply at $T=4$
   MeV and relatively smoothly at $10$ and $16$ MeV as functions of
   density.   

\begin{figure}
\begin{center}
\includegraphics[width=\textwidth]{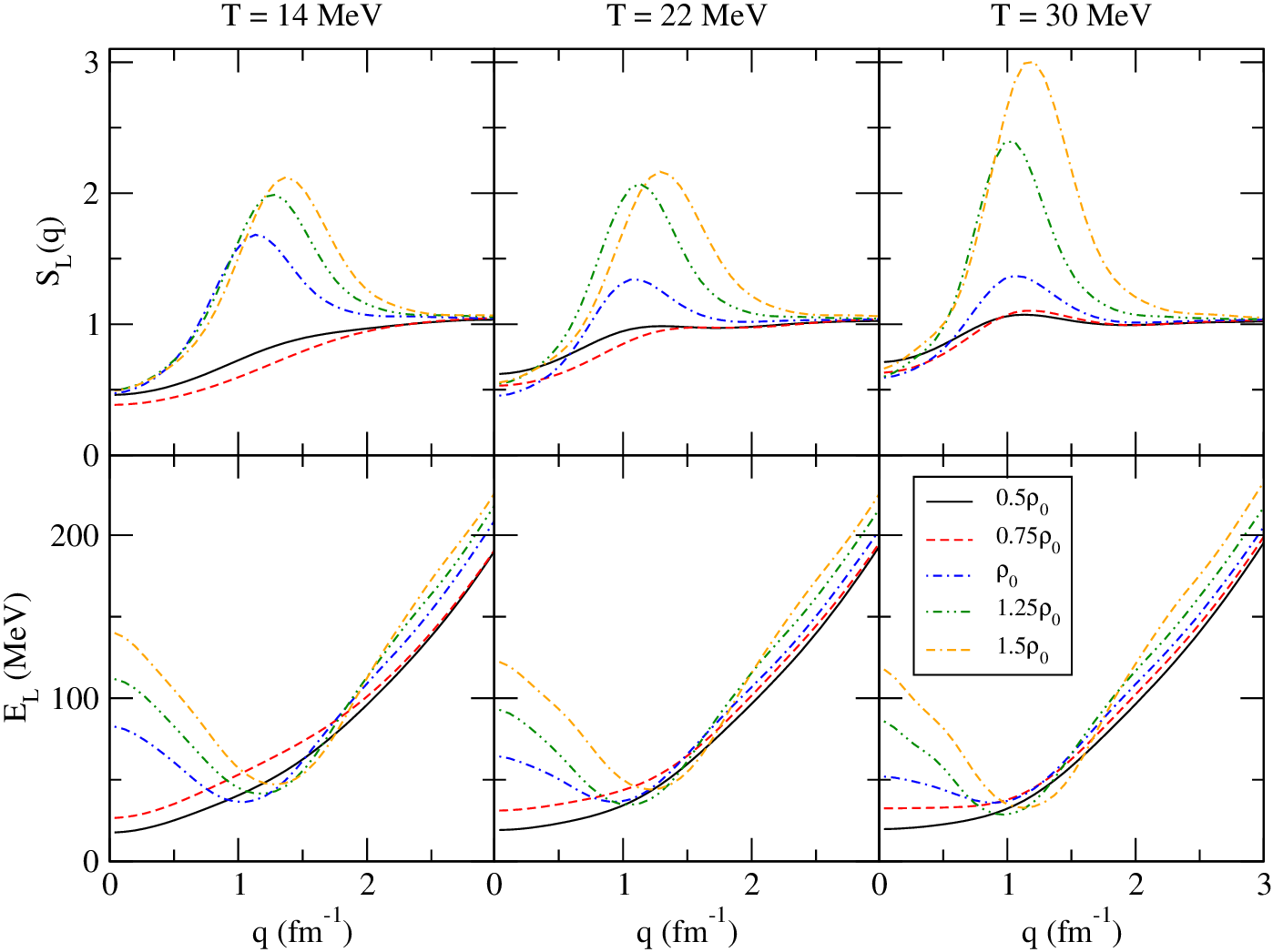}
\caption{ (Color online) The isovector spin longitudinal static structure function and mean energy in pure neutron matter.}
\label{fig:fig7}
\end{center}
\end{figure}

   The same quantities for pure neutron matter at $T = 14$, $22$ and $30$ MeV are
   plotted in Fig.~\ref{fig:fig7}. Again, we see a large enhancement in
   the value $S_L$ for $\rho \ge \rho_0$ at $q \sim 1.3$ fm$^{-1}$ at
   all three temperatures. However the enhancement develops suddenly in the case of
   $T=14$ MeV, and relatively smoothly for $T=22$ and $30$ MeV. Also, the
   mean energy $\bar{E}_L$ shows a dip for densities $\rho \ge \rho_0$ at
   all temperatures, but this effect develops more smoothly at higher temperatures.    
 
   Our calculations thus show that the isovector spin longitudinal response for symmetric nuclear matter is enhanced
   and softened at densities $\rho \gtrsim 2 \rho_0$ at finite temperatures. But whereas for $T<10$ MeV this enhancement and softening develops quite suddenly, there is a smoother evolution for $T \geq 10$ MeV. For pure neutron matter there is enhancement and softening of the isovector spin longitudinal response at densities  $\rho \gtrsim \rho_0$, but it develops quite suddenly for  
$T <22$ MeV and quite smoothly for $T \geq 22$ MeV.

\begin{figure}
\begin{center}
\includegraphics[width=\textwidth]{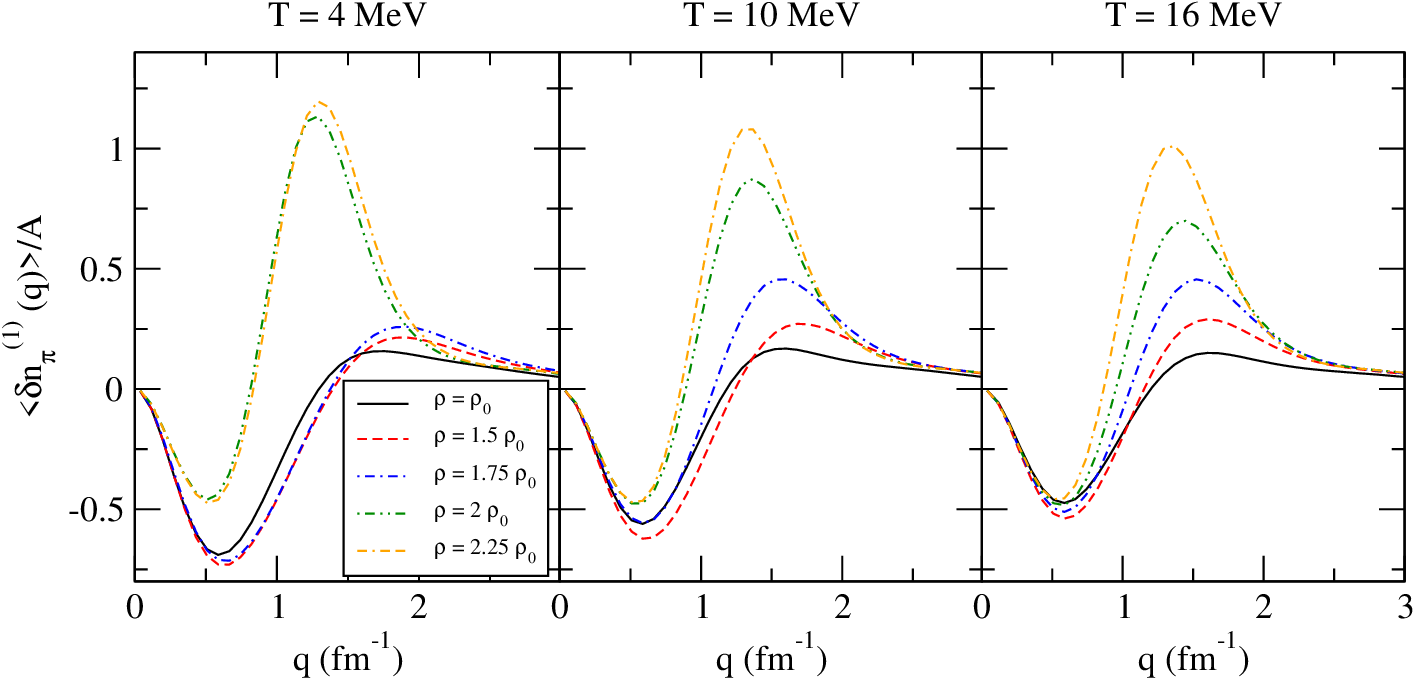}
\caption{ (Color online) The pion excess due to one pion exchange interactions in symmetric nuclear matter.}
\label{fig:fig8}
\end{center}
\end{figure}

\begin{figure}
\begin{center}
\includegraphics[width=\textwidth]{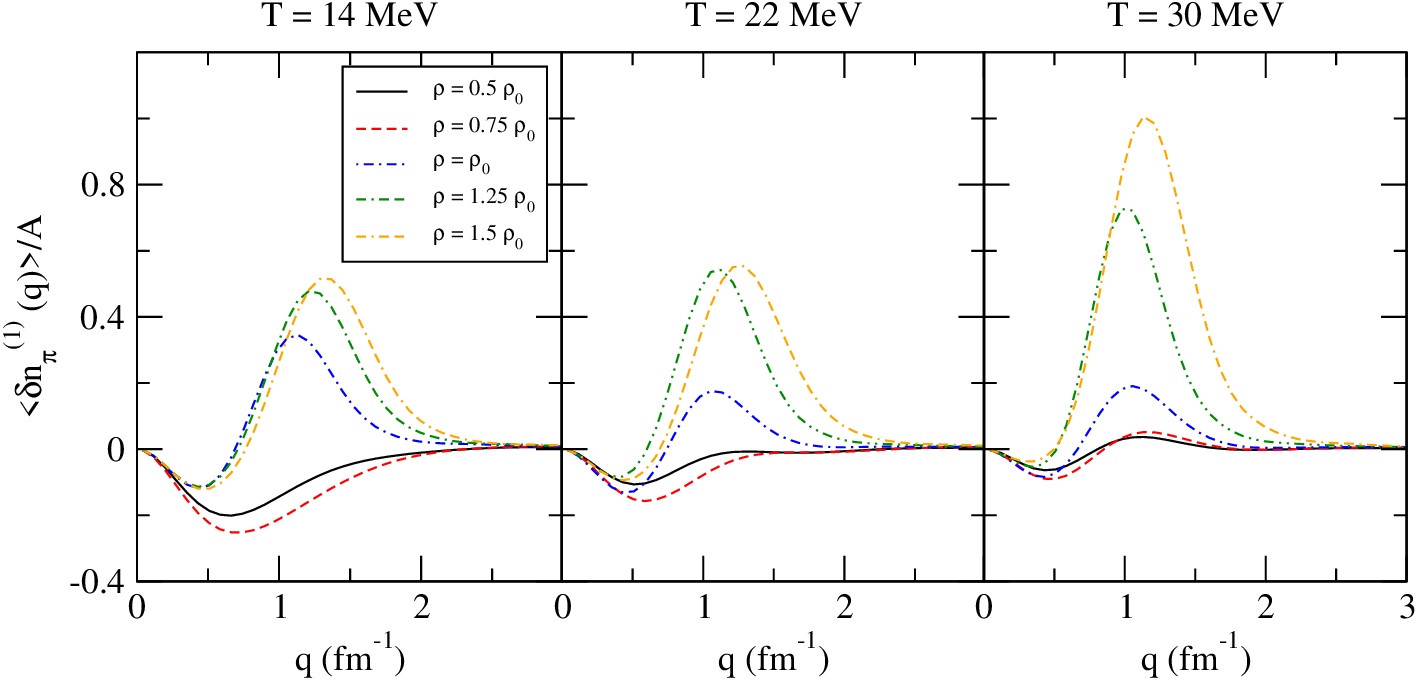}
\caption{ (Color online) The pion excess due to one pion exchange interactions in pure neutron matter.}
\label{fig:fig9}
\end{center}
\end{figure}

   The difference between the expectation values of the pion number operator in a system of $A$ interacting nucleons and in a system of $A$ isolated nucleons is called the pion excess. The part $\langle \delta_{\pi}^{(1)} (q) \rangle$ of the pion excess at a momentum $\bfq$, exclusively due to one pion exchange interactions, can be calculated from $S_L(q)$ \cite{FriPanWir1983, AkmPan1997}. In Figs.~\ref{fig:fig8} and \ref{fig:fig9} we show  $\langle \delta_{\pi}^{(1)} (q) \rangle$ for symmetric nuclear matter and pure neutron matter. As with $S_L(q)$,  $\langle \delta_{\pi}^{(1)} (q) \rangle$ shows an enhancement at higher densities. For symmetric nuclear matter this enhancement develops suddenly for $T < 10$ MeV and smoothly for $T \gtrsim 10$ MeV. In pure neutron matter $T \approx 22$ MeV marks the temperature below which the enhancement is sudden and above which the enhancement is smooth. 
  
   The above observations along with the results for the equation of state (described in Section~\ref{subsec:eos} below) leads us to conclude that the first order phase transition due to neutral pion condensation has a critical temperature of $T_c \approx 10$ MeV for symmetric nuclear matter and $T_c \approx 22$ MeV for pure neutron matter. Above the critical temperature the enhancement in the spin-isospin correlations still persists at higher densities, but now there is a smooth crossover between the low density phase and the high density phase.

\subsection{The Nucleon Effective mass}

\begin{figure}
\begin{center}
\includegraphics[width=\textwidth]{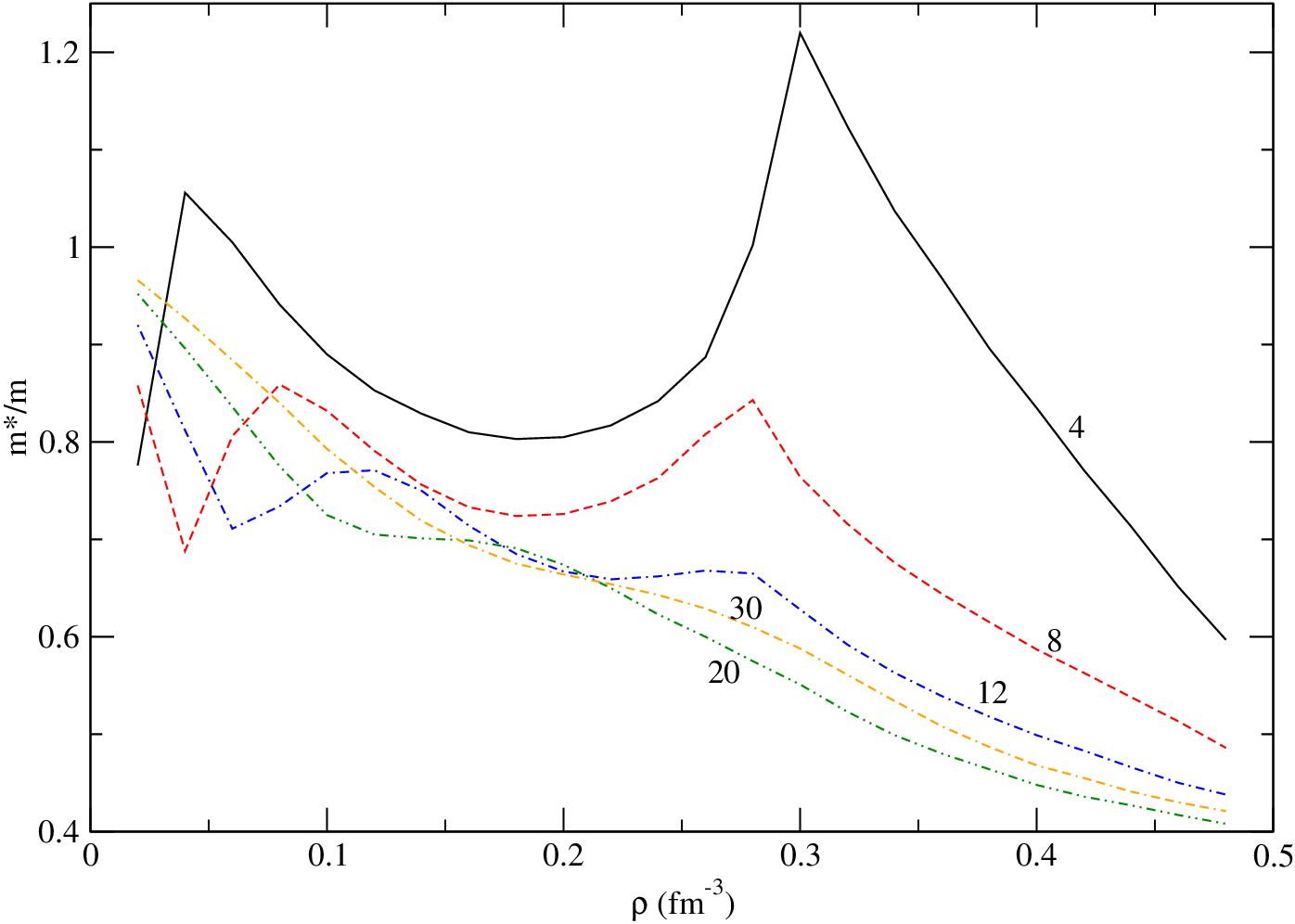}
\caption{ (Color online) The effective mass in symmetric nuclear matter. The numbers alongside the curves denote the temperature in MeV.}
\label{fig:fig10}
\end{center}
\end{figure}

\begin{figure}
\begin{center}
\includegraphics[width=\textwidth]{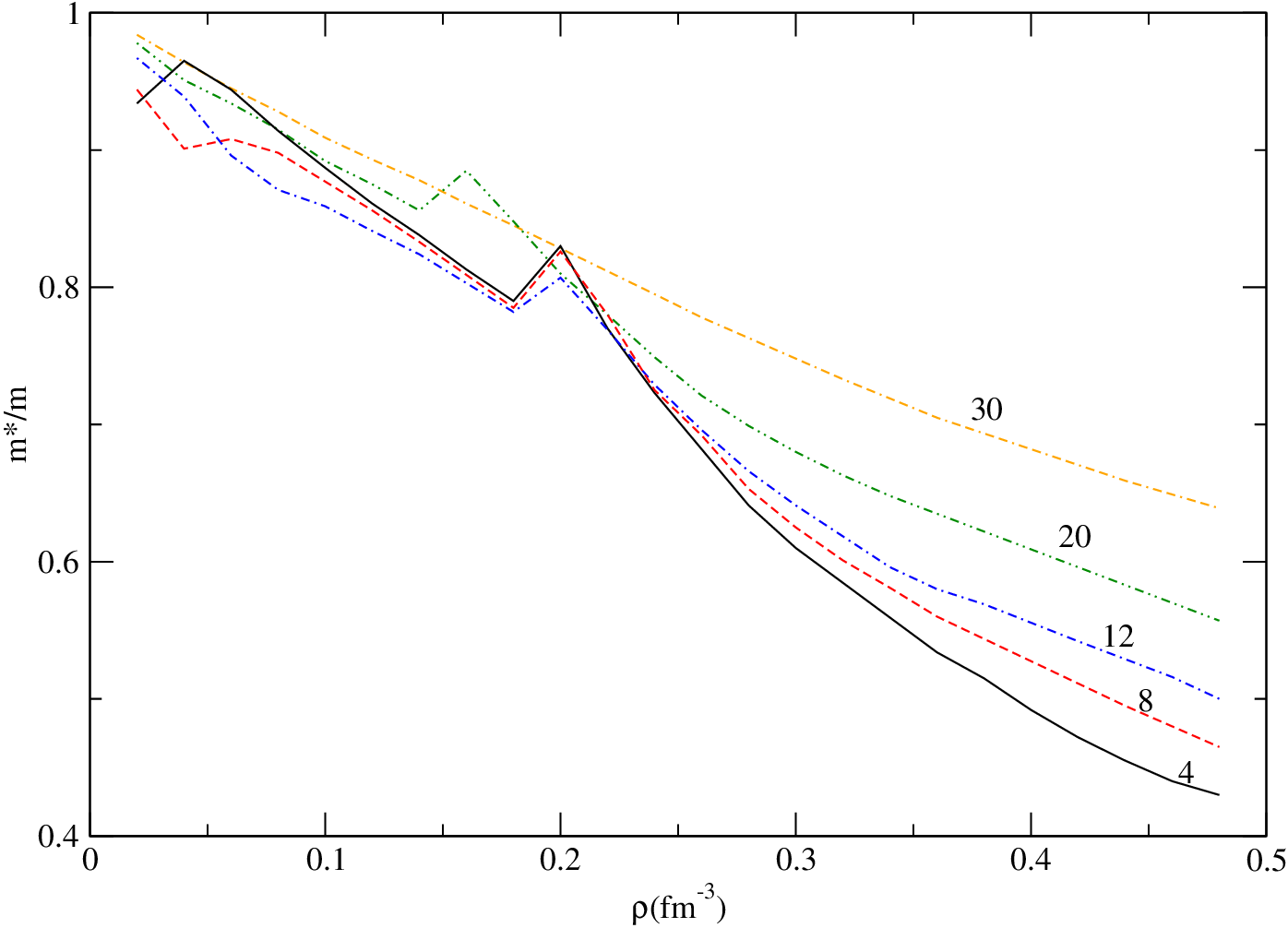}
\caption{ (Color online) The effective mass in pure neutron matter. The numbers alongside the curves denote the temperature in MeV. }
\label{fig:fig11}
\end{center}
\end{figure}

The effective mass $m^{\star}$ is not only a variational parameter in
our calculations, but also a quantity of considerable physical
importance in determining the thermal properties of nuclear
matter. Our results for  $m^{\star}$ as a function of density in symmetric nuclear matter and pure neutron matter for various
temperatures is shown in Figs.~\ref{fig:fig10} and  \ref{fig:fig11}.
 We wish to point out that the variational minimization in pure neutron matter is not very sensitive to
variations of $m^{\star}$, especially at low temperatures and the high density phase. In particular the uncertainty
in our estimate of $m^{\star}$ for the high density phase in pure neutron matter for the three lowest temperatures reported ($4$, $8$ and $12$ MeV)  is comparable to the total temperature dependence of $m^{\star}$ at these temperatures.  

The curves of $m^{\star}$ vs $\rho$ show sharp changes at  $\rho
\sim 2 \rho_0$  in symmetric nuclear matter  and at $\rho \sim \rho_0$ in pure neutron matter. The origin of these changes is
the enhancement of the spin-isospin correlations discussed earlier.

\begin{figure}
\begin{center}
\includegraphics[width=\textwidth]{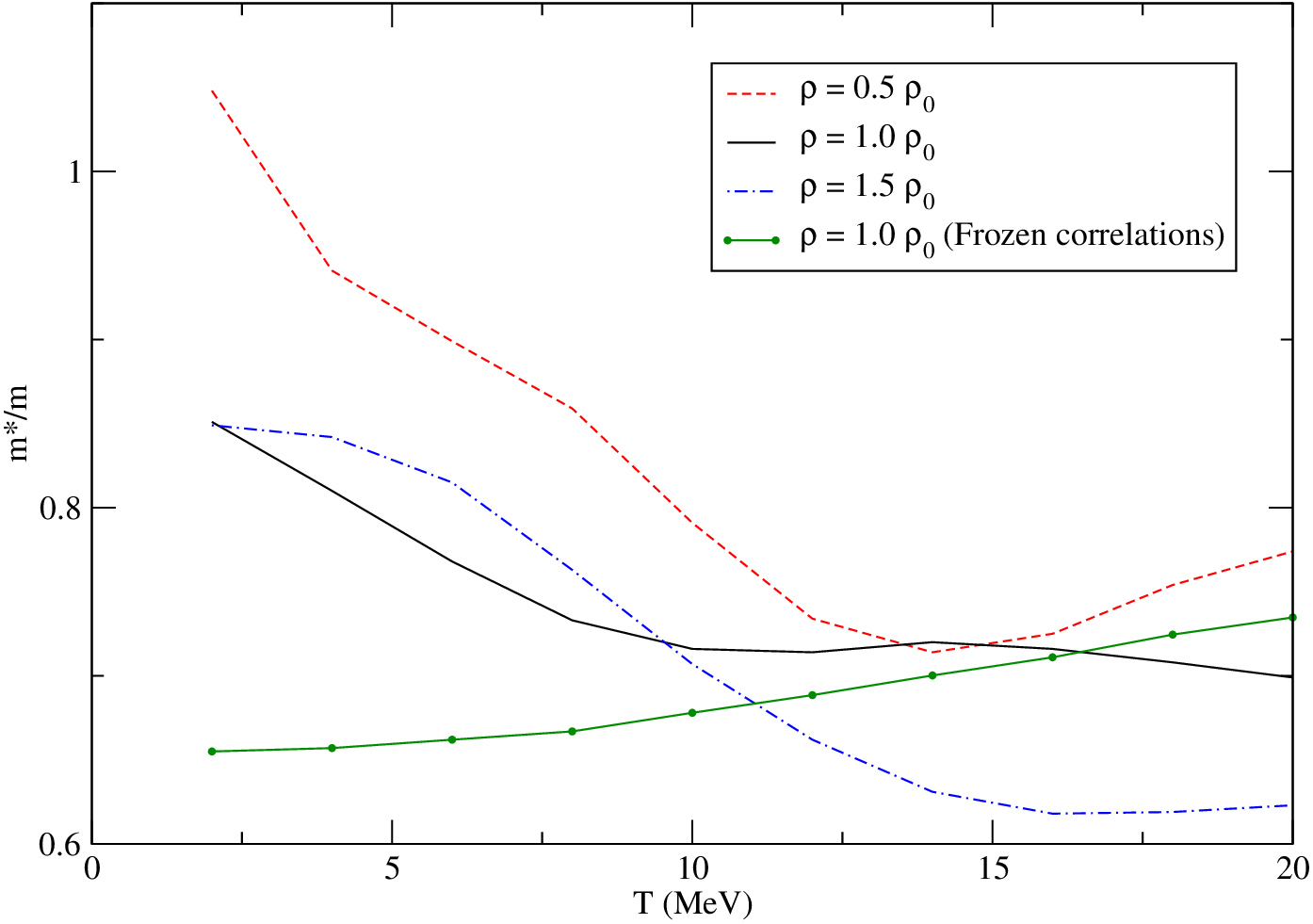}
\caption{ (Color online) The effective mass in symmetric nuclear matter at $\rho = 0.5 \rho_0$, $\rho_0$ and
$1.5 \rho_0$. Also plotted is the effective mass at $\rho=\rho_0$, obtained by using the frozen correlation method described in the text. }
\label{fig:fig12}
\end{center}
\end{figure}

Our calculations also show an enhancement of $m^{\star}$ at low
temperatures in symmetric nuclear matter and to a much lesser extent in pure neutron matter. This is clearer
from Fig.~\ref{fig:fig12} where $m^{\star}$ has been plotted as a
function of temperature for $\rho = 0.5 \rho_0$, $\rho_0$ and $1.5
\rho_0$ in symmetric nuclear matter. For example, for $\rho \sim \rho_0$, $m^{\star}$  at $T=2$ MeV
is $0.85m$ and at $T = 4 $ MeV  is $0.81m$ while at $T=20$ MeV it is $0.69m$
which is very close to the value $0.7m$ obtained from optical
potential models of nucleon-nucleus scattering \cite{BohMot1969,
  JeuLejMah1976}.

 It is important to realize that the effective mass, defined as 
\beq
 m^{\star}(k) = k \frac{\mbox{d}k}{\mbox{d}\epsilon(k)} \ ,
\eeq
can in general depend on the momentum $k$. The momentum
\emph{independent} effective mass $m^{\star}(\rho,T)$ that is used in the variational
calculations is an weighted average of this momentum
\emph{dependent} $m^{\star}(k)$. It is difficult to establish the
actual weighting given to each single particle state. However it is
not difficult to convince oneself that at any given temperature and
density the maximum contribution comes from states  with single particle
energies $\epsilon(k)$ such that 
\beq
\left |\epsilon(k)-\epsilon(k_F) \right | \lesssim \pi T
\eeq        
where $k_F$ is the Fermi momentum. 

At very low temperatures
only those single particle states which are very close to the
Fermi surface can contribute. Theoretical calculations and experimental
evidence suggests that the (momentum \emph{dependent}) $m^{\star} (k)$ in symmetric nuclear matter has a big enhancement at and near the
Fermi surface \cite{JeuLejMah1976, BlaFri1981, KroSmiJac1981,
  FanFriPan1983, BohLanMar1979}. In our calculations 
enhancement in $m^{\star}$ at low temperatures is probably due to this
effect.

 In FP no such enhancement was seen. For comparison, in Fig.~{\ref{fig:fig12}} we have also plotted the $m^{\star}$ at $\rho=\rho_0$ obtained by simply minimizing $F_v$ and using correlation operator from the $T=0$ calculations, i.e. the frozen correlation method. This method is closest to the one used in FP albeit with different nucleon-nucleon interaction and three nucleon interaction. And indeed, we do not see any significant variation of $m^{\star}$ with temperature. In fact, the values of $m^{\star}/m$ so obtained are $0.66$, $0.68$, $0.71$, $0.73$ at $T=5$, $10$, $15$ and $20$ MeV, respectively. These are very close to the corresponding values $0.65$, $0.67$, $0.70$ and $0.73$ obtained by FP where the Urbana $v14$ model of the nucleon-nucleon interaction was used along with a density dependent term to simulate the effects of the three body forces. 

It appears that we are able to capture some of the subtle correlations which influence the low energy
quasiparticle excitations better because of the improvements
introduced in our calculations. These improvements include the fact that in the variational search
all the parameters were varied and the pair
correlation functions were calculated at all temperatures, and the fact that we insist on satisfying the laws of conservation of mass and charge at the level of a few percents. However, this is simply one possibility since in general the free energy is not very sensitive to the variations in the effective mass and  it is difficult to delineate the effect that each individual component in our method has on the effective mass. This topic deserves further investigation, and we hope to address it in more detail in  future work \cite{MukFut}.

  \subsection{The Equation of State}
\label{subsec:eos}
As we discussed earlier the variational minimization for the optimal
parameters $d_t$, $d_c$, $\alpha$ and $m^{\star}$ is carried out with
$v_{18}+V_{\mbox{\tiny{UIX}}}$ only. Following APR some additional corrections were added to the energy.  

\begin{enumerate}
\item The relativistic boost corrections $\delta v_b +
  (V^{\star}_{\mbox{\tiny{UIX}}}-V_{\mbox{\tiny{UIX}}})$ which were
  discussed earlier.
\item An estimate for the second order perturbative corrections
  $\Delta E_2$ is given by $\delta E_{2B}$ the difference between the
  two body cluster energy obtained by minimizing the contributions in each partial
  wave and the two body cluster energy obtained from the $\calF_{ij}$
  defined above.
\item An additional ad-hoc correction term $-\gamma_2 \rho^2
  \mbox{e}^{-\gamma_3 \rho}$, with $\gamma_2=1996$ MeV fm$^6$ and
  $\gamma_3=15.24$ fm$^3$ is added to the symmetric nuclear matter energy to get the correct saturation
  energy at zero temperature. These values are slightly different from
  those in APR ($\gamma_2=2822$ MeV fm$^6$ and $\gamma_3=18.34$ fm$^3$) because our variational energies and variational parameters at zero temperature  are  slightly different from those in APR owing to the improvements in  the method described earlier.
\end{enumerate}

  Of the above only the relativistic boost corrections have a   temperature dependence. The temperature dependence of the perturbative
  corrections was found to be negligible and was neglected.
  Thus our best estimate for the free energy is given by 
\ba
  F_{\mbox{\tiny{tot}}}&=&E_v-T S_v +E_c \\
  &=&F_v+E_c \label{ftot} 
\ea 
where $E_c$ is the sum of the correction terms
  mentioned above.

\begin{figure}
\begin{center}
\includegraphics[width=\textwidth]{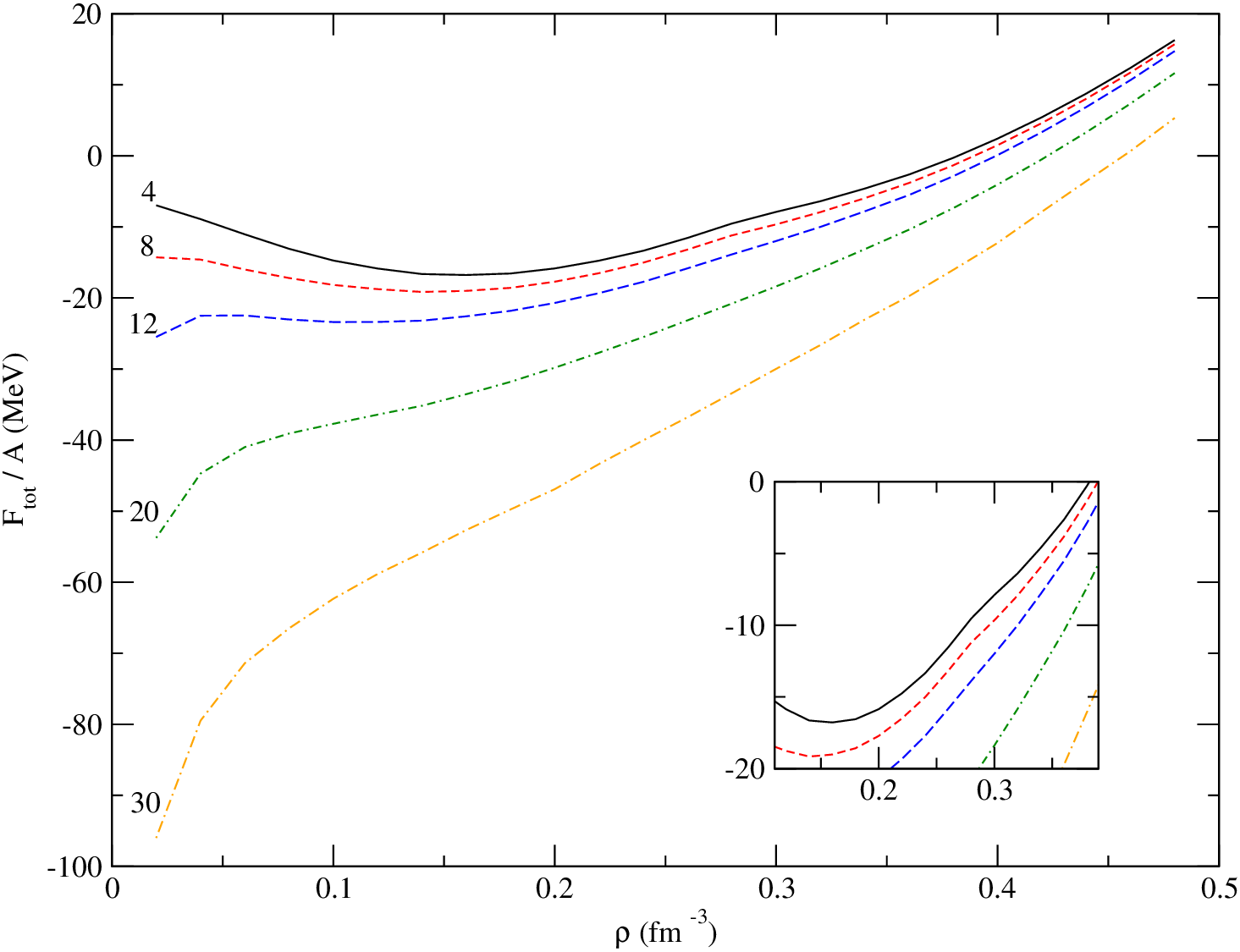}
\caption{ (Color online) The free energy in symmetric nuclear matter. The numbers alongside the curves denote the temperature in MeV. The inset shows in more detail the region of transition from the low density phase to the high density phase.  }
\label{fig:fig13}
\end{center}
\end{figure}

\begin{figure}
\begin{center}
\includegraphics[width=\textwidth]{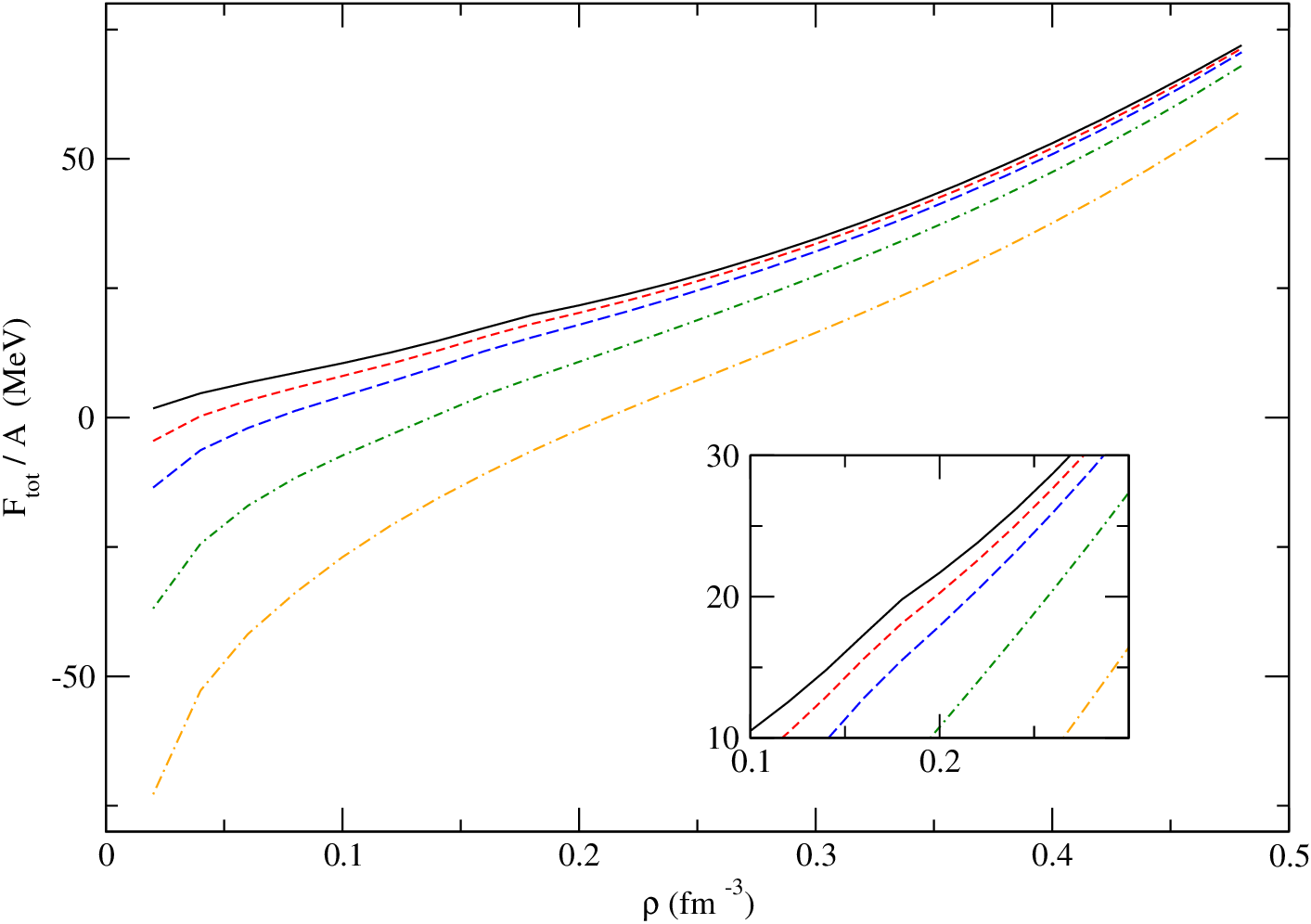}
\caption{ (Color online) The free energy in pure neutron matter. The numbers alongside the curves denote the temperature in MeV. The inset shows in more detail the region of transition from the low density phase to the high density phase.}
\label{fig:fig14}
\end{center}
\end{figure}

  The free energy $F_{\mbox{\tiny{tot}}}$ calculated using the method outlined in the
  last section is shown in Figs.~\ref{fig:fig13} and \ref{fig:fig14} for symmetric nuclear matter and
  pure neutron matter, respectively, for temperatures $T=4-30 $ MeV.  The most striking
  feature in these figures is the sharp change in the slope of the
  free energy at low temperatures for both symmetric nuclear matter ($T<12$ MeV, $\rho \sim
  0.30$ fm$^{-3}$) and pure neutron matter ($T<24$ MeV and $\rho \sim 0.18$
  fm$^{-3}$) (see inset in Figs.~\ref{fig:fig13} and \ref{fig:fig14}). 

\begin{figure}
\begin{center}
\includegraphics[width=\textwidth]{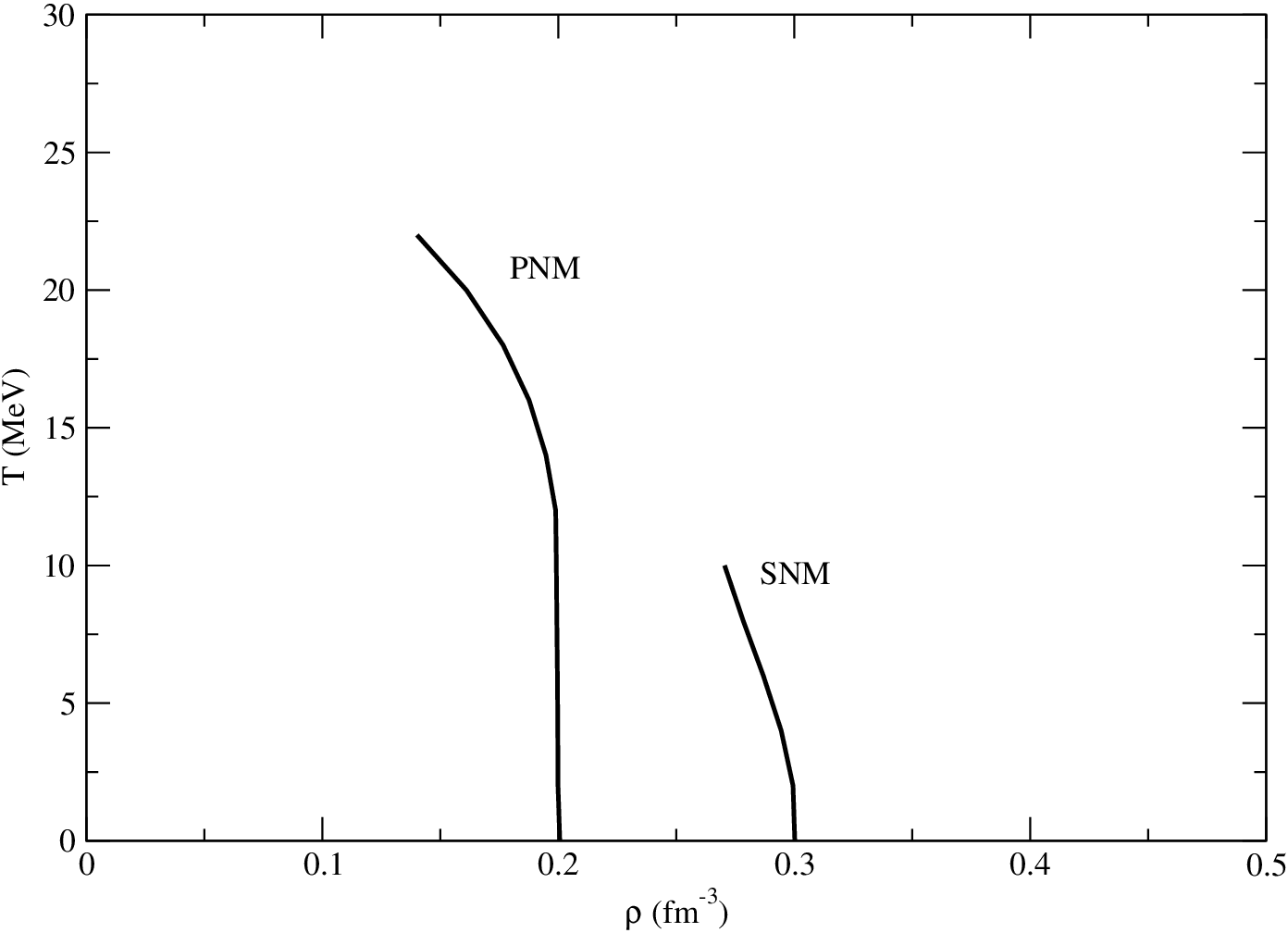}
\caption{ The transition density for neutral pion condensation for symmetric nuclear matter and pure neutron matter in the variational calculations.}
\label{fig:fig15}
\end{center}
\end{figure}

  During the variational search the global minimum of the
  free energy as a function of the variational parameters jumps from
  one local minimum to another. In Fig.~\ref{fig:fig15} we show the density at which this transition from the  low density phase  to high density phase occurs in our calculations. This results in a first order phase transition. In light of our discussion in Section~\ref{subsec:den}, we identify this phase transition with the 
  phenomenon of the neutral pion condensation.  As we mentioned in the earlier the critical temperature for this transition is
  $T_c \approx 10 $ MeV  for symmetric nuclear matter and $T_c \approx 22$ MeV for pure neutron matter.

  At subsaturation densities and low temperatures symmetric nuclear matter undergoes
  another first order phase transition, the liquid-gas
  transition. In  our calculations the  critical temperature for the
  liquid-gas phase transition is $T_c \approx   21$ MeV and the critical density 
  is about $\rho_c \sim 0.3
  \rho_0$. If the calculations are done using the variational free energy $F_v$ only, then the critical temperature for the liquid gas phase transition comes out to be about $16$ MeV. In the
  calculation due to FP the critical temperature is about $17.5$
  MeV. 

The pressure can be calculated from,
\beq
 P(\rho,T) = \rho^2 \left . \frac{\partial F/A}{\partial \rho} \right |_{T} \ .  
\eeq

\begin{figure}
\begin{center}
\includegraphics[width=\textwidth]{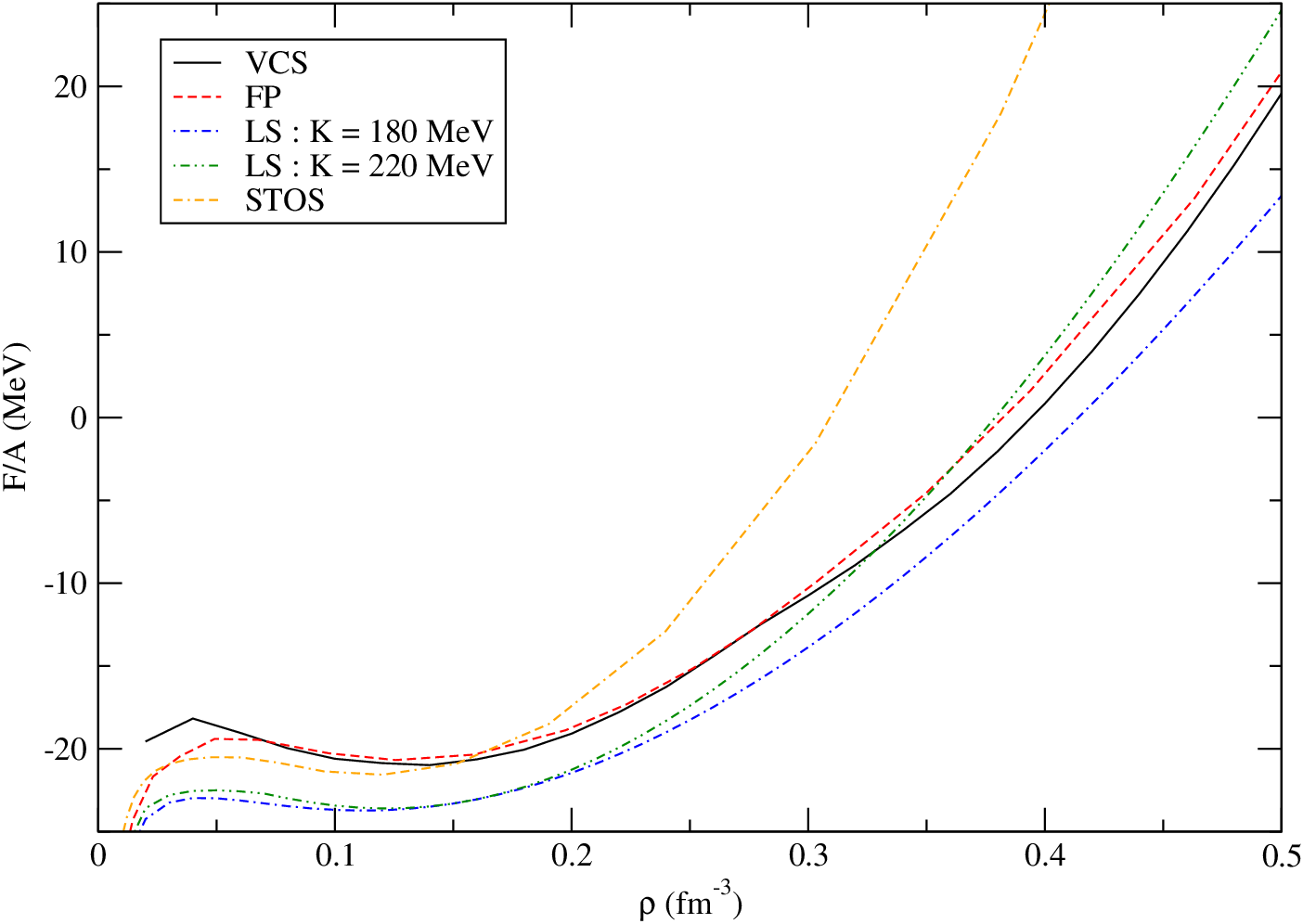}
\caption{ (Color online) A comparison of the free energy obtained
  using various methods at $T=10$ MeV. For an explanation of the
  various curves see the text.}
\label{fig:fig16}
\end{center}
\end{figure}

\begin{figure}
\begin{center}
\includegraphics[width=\textwidth]{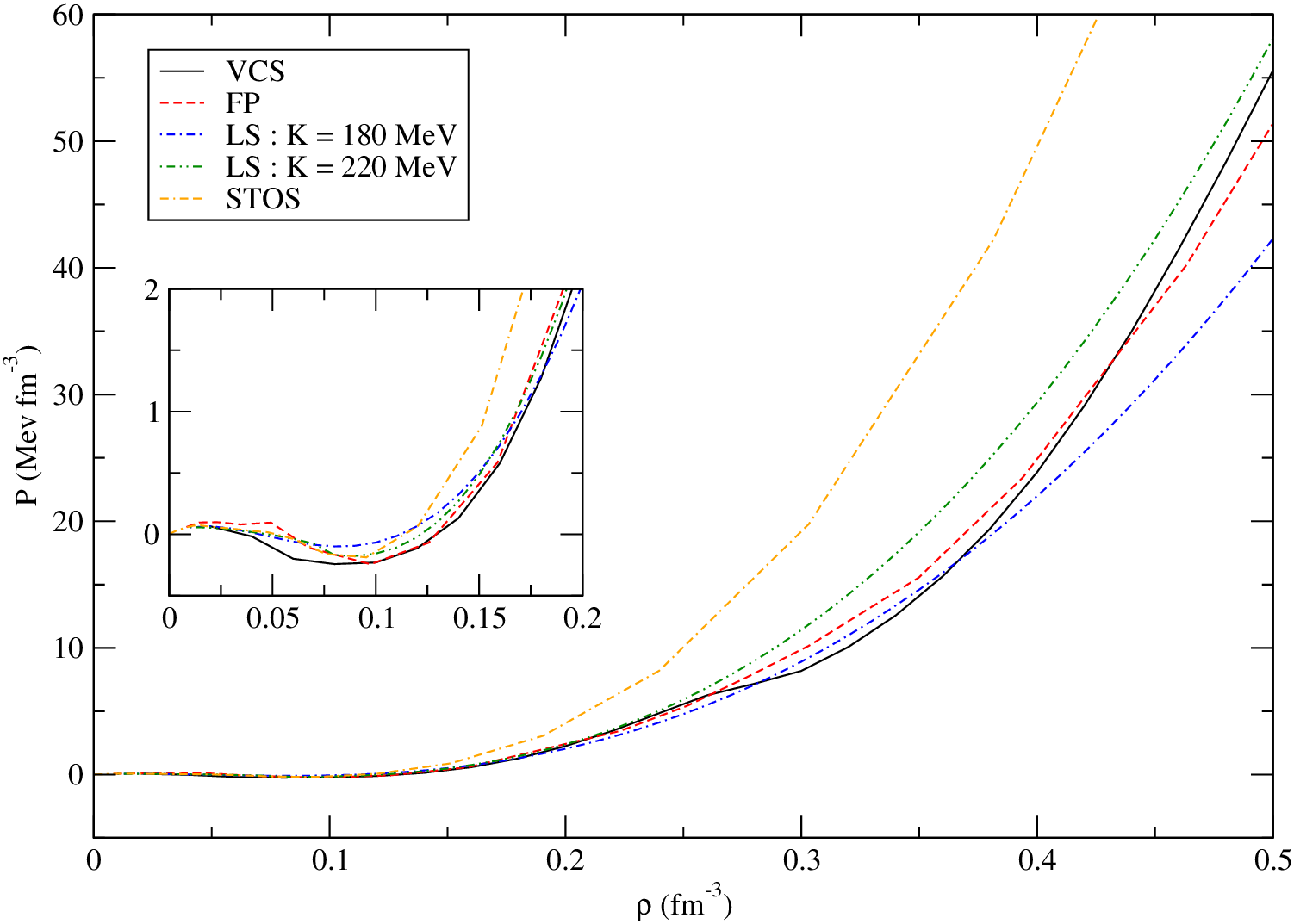}
\caption{ (Color online) A comparison of the pressure obtained
  using various methods at $T=10$ MeV. The inset shows the low density region in more detail. For an explanation of the
  various curves see the text.}
\label{fig:fig17}
\end{center}
\end{figure}

    Our estimates for the free energy  $F(\rho,T)$ ($F_{\mbox{\tiny{tot}}}$ of Eq.~(\ref{ftot})) and the pressure $P(\rho,T)$ of symmetric nuclear matter at $T=10$ MeV from the variational chain summation
  calculations have been plotted in Figs.~\ref{fig:fig16} and \ref{fig:fig17} along with the same from
  some of the other popular equations of state viz. the equation of state due to FP, the Lattimer-Swesty (LS)
  liquid droplet model calculations with incompressibility $K=180$ and
  $220$ MeV \cite{LatSwe1991}
  and the equation of state due to Shen \emph{et al.} (STOS) 
  \cite{She1998npa, She1998ptp} which is based on relativistic mean
  field theory. The inset shows the pressure in the low density region
  in greater detail. We have kept the thermodynamically unstable
  region in the $P$ vs $\rho$ curve to facilitate comparison with the
  other equations of state. Due to the enhanced spin-isospin correlations our equation of state is
  relatively soft at $\rho \sim 2\rho_0$ fm$^{-3}$ but  hardens at
  higher densities.
  
  Various astrophysical phenomena depend sensitively on the symmetry
  energy $E_{\mbox{\tiny{sym}}}$ \cite{LatPra2000} and the symmetry free energy
  $F_{\mbox{\tiny{sym}}}$ \cite{Xu2007} which are defined as 
  \ba
   E_{\mbox{\tiny{sym}}} &=& \frac{1}{2} \left . \frac{\mbox{d}^2 E}{\mbox{d}
       \delta^2} \right |_{\delta =0}   \\
   F_{\mbox{\tiny{sym}}} &=& \frac{1}{2} \left . \frac{\mbox{d}^2 F}{\mbox{d}
       \delta^2} \right |_{\delta=0}
   \ . 
  \ea
 Here $\delta$ is the asymmetry parameter given  by 
 \beq
 \delta=\frac{\rho_n-\rho_p}{\rho},
 \eeq 
where $\rho_n$ is the neutron density and $\rho_p$  is the proton density. To the
 leading order in $\delta$, $E_{\mbox{\tiny{sym}}}$ is
 given by the difference between the energy  in pure neutron matter
 $E_{\mbox{\tiny{PNM}}}$  and the energy  in
 symmetric nuclear matter  $E_{\mbox{\tiny{SNM}}}$. Similarly $F_{\mbox{\tiny{sym}}}$ is given by the difference between the free energy in  pure neutron matter $F_{\mbox{\tiny{PNM}}}$ and symmetric nuclear matter $F_{\mbox{\tiny{SNM}}}$.
   \ba
   E_{\mbox{\tiny{sym}}} &=& E_{\mbox{\tiny{PNM}}} - E_{\mbox{\tiny{SNM}}} + \calO(\delta^2)   \\
   F_{\mbox{\tiny{sym}}} &=& F_{\mbox{\tiny{PNM}}} - F_{\mbox{\tiny{SNM}}} + \calO(\delta^2)   \ . 
  \ea
  In Figs.~\ref{fig:fig17} and \ref{fig:fig19} we show $E_{\mbox{\tiny{PNM}}} - E_{\mbox{\tiny{SNM}}}$ and
  $F_{\mbox{\tiny{PNM}}} - F_{\mbox{\tiny{SNM}}}$  as functions of
  density at various temperatures.

\section{Discussion}
\label{sec:dis}

In this section we will discuss the limitations
of our calculations. First, let us consider the validity of the
quasiparticle picture at finite temperature. This assumption
is motivated by Landau's theory of Fermi liquids. However Landau's
theory in its original formulation applies only to systems at zero or
very low temperatures. It is justified on the basis of the fact that
low lying quasiparticle excitations (as defined by the poles in the single
particle propagator) are long lived at low temperatures
\cite{BayPet1978flt, AbrKha1959}. At higher
temperatures this is not true and one cannot associate well defined
quasiparticle excitations to the system.   

It is, however, important to recognize that there are two separate
issues involved here. On the one hand, we have the thermodynamical statement that the free energy of the
system can be written as a functional of quasiparticle occupation
numbers and that in particular the entropy is given by Eq.~(\ref{entpy}). On the other hand,  we have one
particular microscopic justification of the previous statement in terms of the long lived
nature of the quasiparticle excitations. It has been shown in the past
that although the latter is true only for very low temperatures, the
former, under very general conditions, is true for arbitrary
temperatures \cite{BalDom1971, Dom1960, Lut1968, PetCar1973,
  PetCar1975}. With this in mind we think that our final estimate for 
the free energy is more reliable than the individual components that
entered the calculations, e.g. the assumption of an one-to-one
mapping between the states of a non interacting system and an
interacting one is not exactly true at for the states relevant at any finite
temperature. However, even with this caveat in mind, we believe that our
predictions about the microscopic structure of the system are useful in
accounting for at least the gross features if not the details in the
many body system.

Our description of the high density phase is clearly
incomplete. Our wave function does not include any true long range
order. It has been shown from very general arguments by Landau and
Peierls that  long range order where the order parameter varies only in one dimension is unstable at
finite temperatures \cite{BayFriGri1982}. However, condensates that vary in two or three dimensions can exist at any temperature. If the high density phase is indeed a state with broken symmetry then it is
possible that the first order transition changes into a second order
transition  beyond  critical temperature (instead of a smooth crossover from the low density phase to the high density phase seen in our calculations). From our calculations we cannot make 
any prediction about this possibility. 
    
In the low density and low temperature region for symmetric nuclear matter the uniform
fluid is no longer the optimal configuration and few body
clusters are expected to emerge (see e.g. \cite{rop2001}). We see the precursor to deuteron
clustering in our calculations, e.g. in Fig.~\ref{fig:fig20} we see that the pair
distribution function for symmetric nuclear matter at $\rho=0.02$ fm$^{-3}$ and $T=2$ MeV
shows a pronounced bump at $r\sim 1$ fm. Genuine clustering of three or more particles, however, is not possible in our
variational chain summation calculations because our wave functions are too simple to describe such phenomena. 

Another phenomenon which  occurs at very low densities and
temperatures in nuclear matter is superfluidity due to the formation
of Cooper pairs \cite{DeaHjo2003, SedCla2006}. Pairing, especially in very low
density pure neutron matter and neutron star matter,  has
been the subject of extensive research in the recent past
\cite{LomSch2001} (and references therein). The
wave functions that we have used do not include the effects of pairing
in the long range part. However we would like to point out that it is
possible to develop a variational theory based on correlated basis states for superfluid
systems at zero temperatures \cite{Fan1981cbcs, Fab2008cbcs}. It is quite possible that this theory can be
extended to finite temperatures in a manner very similar to the one
used here for normal systems.

At very high temperatures ($T\gtrsim 50$ MeV for symmetric nuclear matter) thermal excitations of
the pionic degrees of freedom, the `thermal pions' \cite{KolBay1982}  become
important and should be explicitly included in the equation of state \cite{FriPanUsm1981} . Our
present calculations do not have any explicit non-nucleonic degrees of
freedom, thus they cannot be extended to very high temperatures without the explicit inclusion of the thermal pions. The
calculation of the pionic spectrum in nuclear matter is a 
challenging task. However, as the first step, the simplified
quasiparticle treatment of the pions along with the full variational chain summation
calculation for the nucleons as done in Ref.~\cite{FriPanUsm1981} might prove useful.

Finally, we would briefly discuss the case of arbitrary proton
fraction. Pure variational chain summation can only be used to calculate the equation of state of symmetric nuclear matter and
pure neutron matter, and not for asymmetric nuclear matter. This problem
can be circumvented by using the so called quadratic approximation,        
where one assumes that the interaction energy at an arbitrary asymmetry $\delta$
has a purely  quadratic dependence on $\delta$.
 Using this fact the energy for any $\delta$ can be found from the results at
$\delta=0$ (symmetric nuclear matter ) and $\delta=1$ (pure neutron matter) \cite{AkmPanRav1998, PanRav1989}.

It has been shown in Ref.~\cite{LagPan1981anm} that within the scheme of variational calculations (although with a lower order approximation for the non-central correlations as compared to the variational chain summation method) the quadratic approximation is quite accurate, at least up to the leading order terms in the interaction energy. This calculation was done for the zero temperature low density phase, with the Urbana $v_{14}$ model for nucleon-nucleon interaction and a density dependent
model for three nucleon interaction. Thus far, the  calculation has not been repeated for finite
temperatures, the high density phase or for the modern sets of nucleon-nucleon interaction and three nucleon interaction. It is, however, plausible that the result
is true for the aforementioned conditions as well.

\section{Conclusion}
\label{sec:con}

The calculation of the properties of nuclear matter in supernovae,
neutron stars and heavy ion collisions starting from the properties
of nucleons in vacuum and light nuclei is an outstanding problem in
nuclear many body theory. In a preceding paper (I) we showed that
when the variational theory is extended to finite temperatures, the
free energy and the single particle energies do not have any
corrections due to the non-orthogonality of the correlated basis states used in these
variational theories. In this paper we showed how this formalism can be used for practical calculations in hot dense nucleon matter. 

We calculated the equation of state of symmetric nuclear matter and pure neutron matter and discussed the fate of the neutral pion condensation at finite temperature. The first order phase transition was seen to have a critical temperature of about $10$ MeV for symmetric nuclear matter and about $22$ MeV for pure neutron matter. However, the the enhancement of the spin-isospin correlations and as a result the softening of the equation of state, near $\rho \sim 2 \rho_0$  for symmetric nuclear matter  and near $\rho\sim \rho_0$ for pure neutron matter, persists even beyond the critical temperature. We also discussed the behavior of the nucleon effective mass and saw an enhancement at lower temperatures. We proposed one possible explanation for this effect. 

The pair correlation functions $\calF_{ij}$ generated during our variational minimizations can be used to calculate the equation of state of asymmetric nuclear matter within the quadratic approximation \cite{PanRav1989}, the single particle energies in dense nuclear matter \cite{FriPan1981sps, Wir1988sps}, effective interactions \cite{CowPan2003, CowPan2006} and transport properties \cite{BenVal2007}. Some of these topics will be explored in our future work \cite{MukFut}.

\section*{Acknowledgements}

 The initial
ideas which led to this work were introduced to the author by
the late V.R. Pandharipande. The author
wishes to thank D. G. Ravenhall for his patient guidance throughout the
length of this work, J. Morales for his help with zero temperature computer program, and G. Baym and R. B. Wiringa for numerous
helpful discussions and suggestions.  This work was supported in part by the National Science Foundation Grant PHY 07-01611.

%\bibliography{refs02}

\newpage

\clearpage    

\clearpage

\clearpage
%two column

\clearpage

\clearpage

\clearpage
%two column

\clearpage
%two column

\clearpage
%two column

\clearpage
%two column

\clearpage

\clearpage

\clearpage

\clearpage

\clearpage

\clearpage

\clearpage

\clearpage

\clearpage
\begin{figure}
\begin{center}
\includegraphics[width=\textwidth]{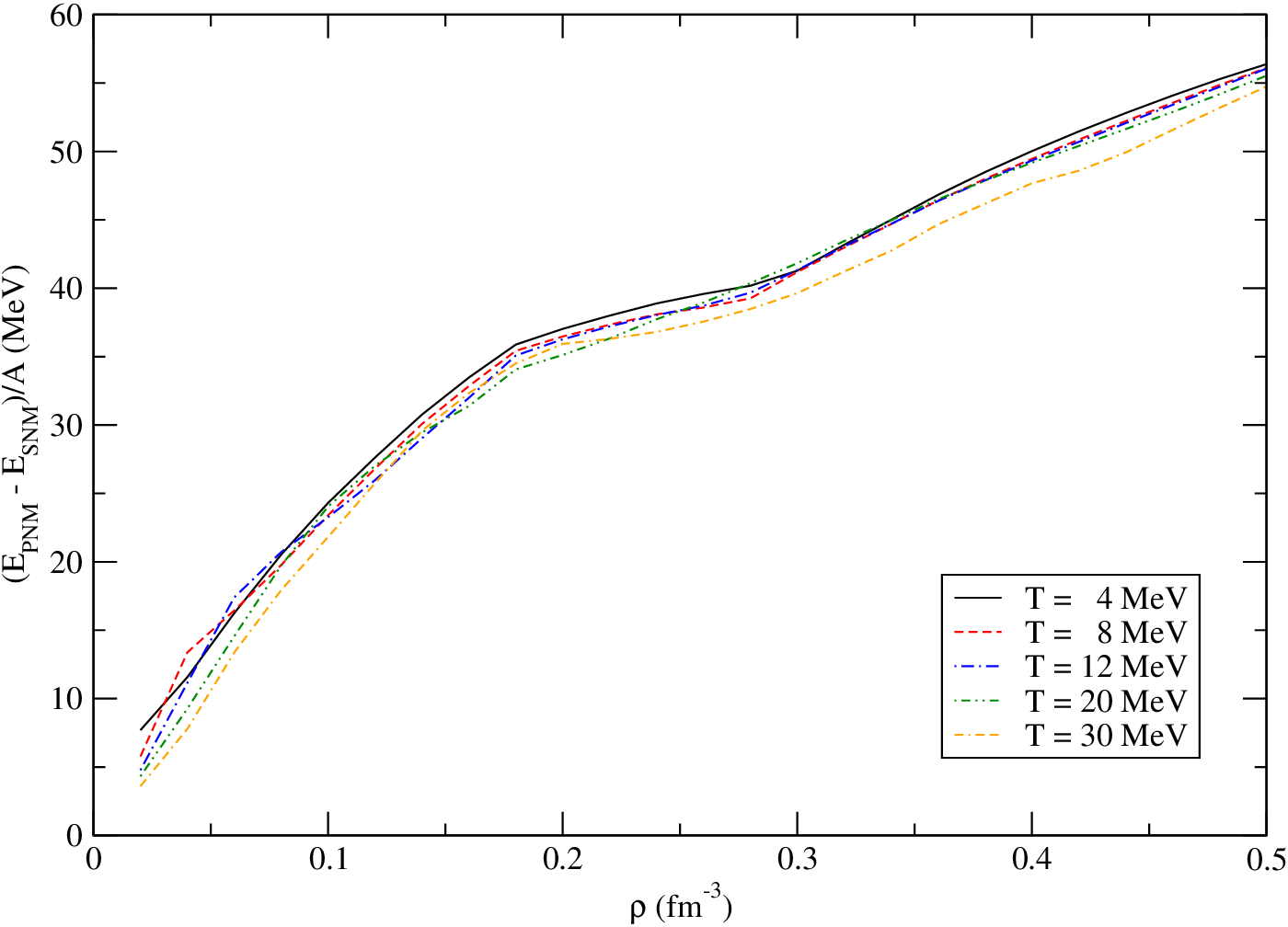}
\caption{ (Color online) A comparison of  $E_{\mbox{\tiny{PNM}}}-E_{\mbox{\tiny{SNM}}}$
  at various temperatures} 
\label{fig:fig18}
\end{center}
\end{figure}

\clearpage
\begin{figure}
\begin{center}
\includegraphics[width=\textwidth]{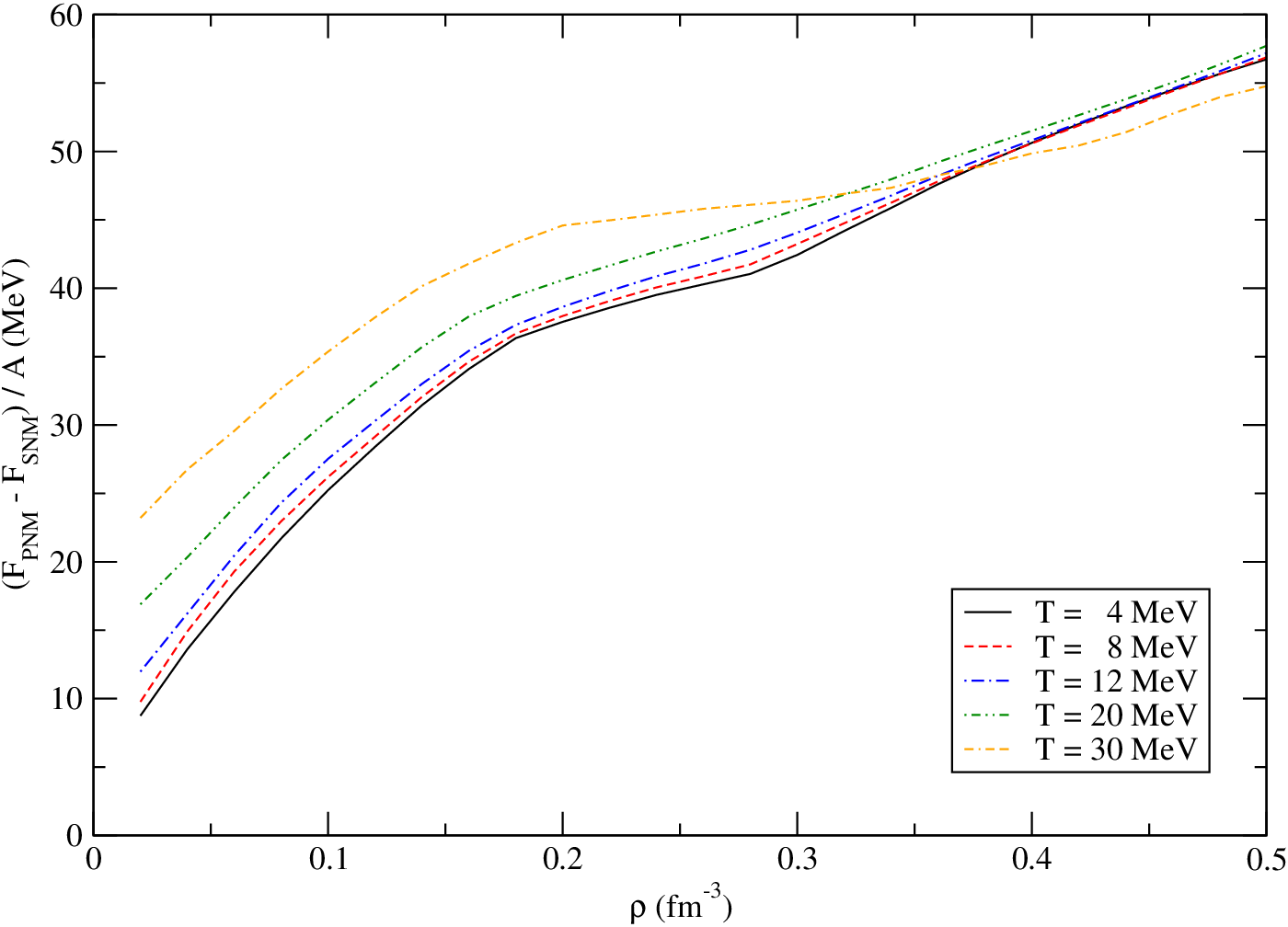}
\end{center}
\caption{ (Color online) A comparison of $F_{\mbox{\tiny{PNM}}}-F_{\mbox{\tiny{SNM}}}$
  at various temperatures} 
\label{fig:fig19}
\end{figure}

\clearpage
\begin{figure}
\begin{center}
\includegraphics[width=\textwidth]{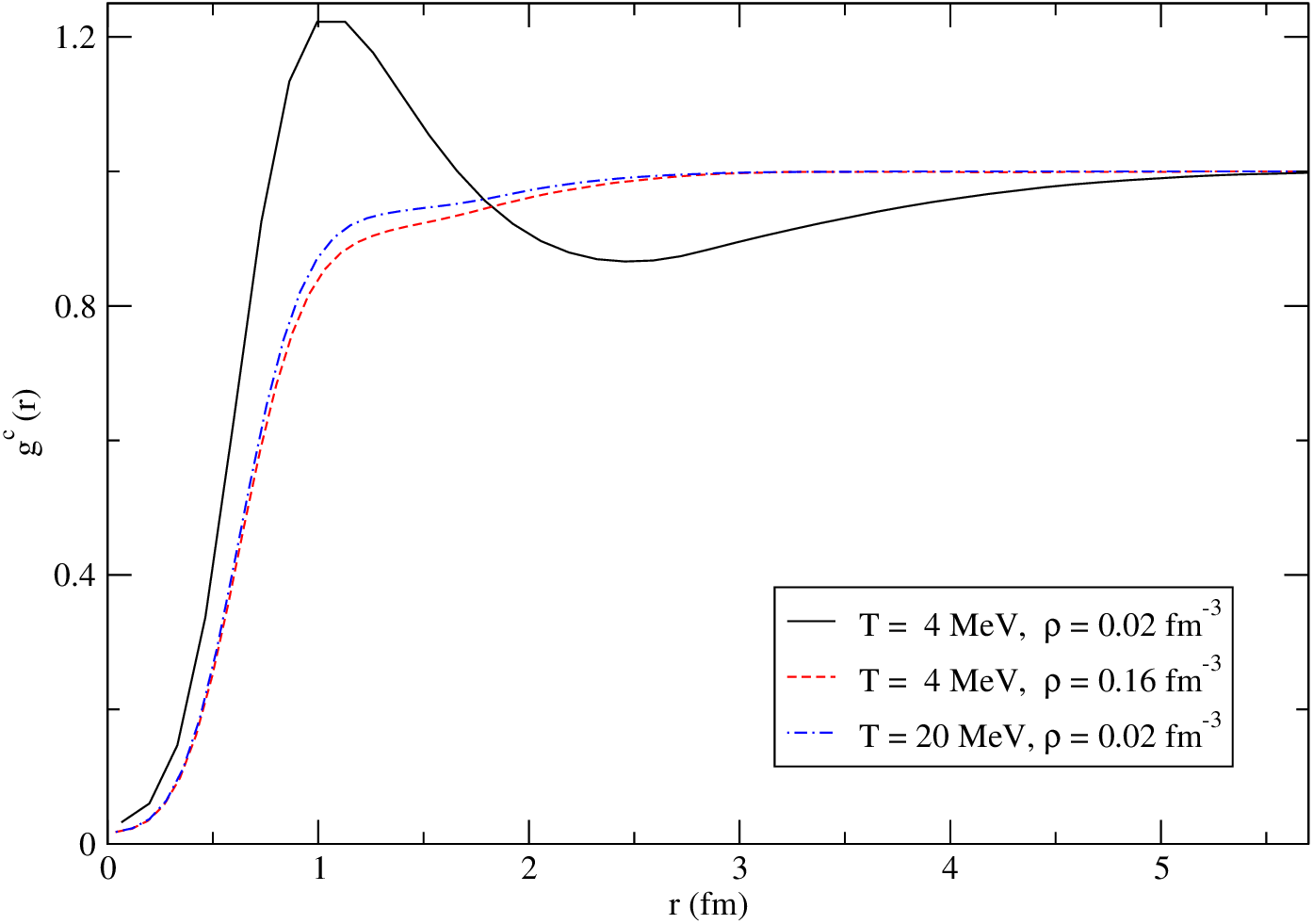}
\caption{ (Color online) The normalized pair distribution function at $T = 4$ MeV and
  $\rho=0.02$ fm$^{-3}$ shows the effects of deuteron clustering. The
  same function has been plotted for $T=20$ and $\rho=0.02$ fm$^{-3}$ and $T=4$ MeV, and
$\rho=0.16$ fm$^{-3}$ for comparison.}
\label{fig:fig20}
\end{center}
\end{figure}

\clearpage

\end{document}